\documentclass{mn2e}
\usepackage{epsfig}
\usepackage{longtable}
\usepackage{graphics}
\usepackage{amssymb}
\usepackage{times}
\title[Photoionization model of 30 Doradus]
{A photoionization modeling study of 30 Doradus: the case for small-scale
chemical inhomogeneity}
\author[Y. G. Tsamis and D. P\'{e}quignot]
{Yiannis G. Tsamis and Daniel P\'{e}quignot\\
LUTH, Laboratoire l'Univers et ses Th\'eories, associ\'e au CNRS (FRE 2462) et
\'a l'Universit\'e Paris 7, Observatoire de Paris-Meudon,\\
F-92195 Meudon C\'edex, France;\\ YGT: P. Gruber Foundation/IAU Fellow;\\
Current address: Dept. of Physics and Astronomy, University College London,
Gower Street, London WC1E 6BT, U.K.; E-mail: ygt@star.ucl.ac.uk}

\date{Received:}
\newcommand{\apj}{ApJ}
\newcommand{\apjs}{ApJS}
\newcommand{\aap}{A\&A}
\newcommand{\aj}{AJ}
\newcommand{\apjl}{ApJL}
\newcommand{\mnras}{MNRAS}
\newcommand{\eld}{$N_{\rm e}$}
\newcommand{\hyd}{$N{\rm (H^{+})}$}

\newcommand{\elt}{$T_{\rm e}$}
\newcommand{\tion}{$T_{\rm 0, i}$}
\newcommand{\teff}{$T_{\rm eff}$}
\newcommand{\hyt}{$T{\rm (H^{+})}$}
\newcommand{\oppt}{$T{\rm (O^{2+})}$}

\newcommand{\cmt}{cm$^{-3}$}
\newcommand{\tsq}{$t^{2}$}
\newcommand{\tsqb}{$t^{2}_{\rm Bal}$}
\newcommand{\tsqd}{$t^{2}_{\textsc{adf}}$}
\newcommand{\cp}{C$^+$}
\newcommand{\cpp}{C$^{2+}$}

\newcommand{\op}{O$^+$}
\newcommand{\nep}{Ne$^+$}
\newcommand{\opp}{O$^{2+}$}

\newcommand{\np}{N$^+$}
\newcommand{\npp}{N$^{2+}$}

\newcommand{\foiii}{[O~{\sc iii}]}
\newcommand{\fniii}{[N~{\sc iii}]}
\newcommand{\foi}{[O~{\sc i}]}
\newcommand{\foii}{[O~{\sc ii}]}
\newcommand{\fsii}{[S~{\sc ii}]}
\newcommand{\fsiii}{[S~{\sc iii}]}
\newcommand{\fsiv}{[S~{\sc iv}]}
\newcommand{\fci}{[C~{\sc i}]}
\newcommand{\fni}{[N~{\sc i}]}
\newcommand{\fnii}{[N~{\sc ii}]}
\newcommand{\farii}{[Ar~{\sc ii}]}
\newcommand{\fariv}{[Ar~{\sc iv}]}
\newcommand{\fcliii}{[Cl~{\sc iii}]}
\newcommand{\fclii}{[Cl~{\sc ii}]}
\newcommand{\fcliv}{[Cl~{\sc iv}]}
\newcommand{\fneii}{[Ne~{\sc ii}]}
\newcommand{\fneiii}{[Ne~{\sc iii}]}

\newcommand{\ffeii}{[Fe~{\sc ii}]}
\newcommand{\ffeiii}{[Fe~{\sc iii}]}
\newcommand{\ffeiv}{[Fe~{\sc iv}]}
\newcommand{\oiii}{O~{\sc iii}}

\newcommand{\nii}{N~{\sc ii}}
\newcommand{\niii}{N~{\sc iii}}

\newcommand{\oi}{O~{\sc i}}
\newcommand{\oii}{O~{\sc ii}}
\newcommand{\cii}{C~{\sc ii}}
\newcommand{\neii}{Ne~{\sc ii}}
\newcommand{\mgii}{Mg~{\sc ii}}
\newcommand{\ciii}{C~{\sc iii}}
\newcommand{\civ}{C~{\sc iv}}
\newcommand{\fciii}{C~{\sc iii}]}

\newcommand{\fariii}{[Ar~{\sc iii}]}

\newcommand{\hi}{H\,{\sc i}}
\newcommand{\hii}{H~{\sc ii}}
\newcommand{\hei}{He~{\sc i}}
\newcommand{\heii}{He~{\sc ii}}

\newcommand{\hp}{H$^+$}
\newcommand{\hep}{He$^+$}

\newcommand{\ha}{H$\alpha$}
\newcommand{\hb}{H$\beta$}
\newcommand{\hg}{H$\gamma$}

\begin{document}
\maketitle

\begin{abstract}

\noindent Photoionization models of the giant \hii\ region 30 Doradus are built
and confronted to available UV, optical, IR (\emph{ISO}) and radio spectra,
under black-body or \emph{CoStar} spectral energy distributions for the primary
source and various density distributions for the nebular gas. Chemically
homogeneous models show very small rms electron temperature fluctuations and
fail to reproduce the heavy element optical recombination line (ORL) spectrum
of the nebula. Dual abundance models incorporating small-scale chemical
inhomogeneities in the form of hydrogen-deficient inclusions which are in
pressure balance with the normal composition ambient gas, provide a better fit
to the observed heavy element ORLs \emph{and} other nebular lines, while most
spectral features are satisfactorily accounted for. The inclusions, whose mass
is $\sim$\,2\,per cent of the total gaseous mass, are 2--3 times cooler and
denser than the ambient nebula. Their O/H abundance ratio is $\sim$\,0.9\,dex
larger than in the normal composition gas and have typical mass fractions of
$X$ $=$ 0.687, $Y$ $=$ 0.273 and $Z$ $=$ 0.040. Helium is found to be about as
deficient as hydrogen in the inclusions, while elements heavier than neon, such
as sulfur and argon, are quite possibly enhanced in proportions similar to
oxygen, as indicated by the most satisfactory dual abundance model obtained.
This suggests that the posited H-deficient inclusions may have arisen from
partial mixing of matter which was nucleosynthetically processed in a supernova
event with gas of normal LMC composition. The average gaseous abundances of the
chemically inhomogeneous models are $\sim$\,0.08\,dex higher than those of the
homogeneous models, yet they are lower by a similar $\sim$\,0.08\,dex than
those derived from standard empirical methods (fully corrected for inaccuracies
in ionization correction factors and differences in atomic data) which
postulate temperature fluctuations in a chemically homogeneous medium.
Attention is drawn to a bias in the determination of \hii\ region helium
abundances in the presence of H-deficient inclusions. It is argued that these
results provide evidence for incomplete small-scale mixing of the interstellar
medium (ISM). The case for the existence of abundance inhomogeneities in \hii\
regions is examined in the light of current theoretical considerations
regarding the process of chemical homogenization in the ISM.

\vspace{0.3cm}

\noindent {\bf Key Words:} ISM: abundances -- {\hii} regions -- ISM: evolution
-- ISM: individual: 30 Dor -- galaxies: individual: LMC

\end{abstract}

\section{Introduction}

The emission-line analysis of ionized gaseous nebulae such as planetary nebulae
(PNe; ionized ejected envelopes of low- to intermediate-mass evolved stars) and
H~{\sc ii} regions (ionized gas clouds marking the birth places of stars, like
the Orion Nebula) is one of the best established methods for obtaining the
abundance of elements in our own and in external galaxies. Abundance studies of
PNe reveal the composition of matter returned to the interstellar medium (ISM)
after being processed by stellar nucleosynthesis, while those of H~{\sc ii}
regions provide a snapshot of the composition of the ISM in the latest episode
of its evolution.

Since before the 1990's, a source of controversy in the field has arisen from
the fact that, for both PNe and H~{\sc ii} regions, the abundances of C, N, O,
and Ne (relative to hydrogen) based on optical recombination lines (ORLs) are
found to be consistently higher than the abundances of the same elements
derived from their UV, optical and infrared collisionally excited lines (CELs;
Barker 1984; Peimbert, Storey \& Torres-Peimbert 1993; Rola and Stasi{\'n}ska
1994; Liu et al. 1995, 2000, 2001, 2004; Esteban et al. 2005; Tsamis et al.
2003a, 2003b, 2004). The ORL/CEL ionic abundance discrepancy factors (ADFs) for
\opp\ and \cpp\ have been shown to be strongly correlated with the difference
between the nebular electron temperatures derived from the nebular to auroral
\foiii\ optical-line ratio and the \hi\ Balmer recombination continuum break at
3646-\AA\ (Liu et al. 2001; Tsamis 2002; Tsamis et al. 2004).

Deciphering the thermal and density structure of ionized nebulae is a very
important matter in its own right, given the exponential dependence of
CEL-based abundances on the adopted nebular electron temperature (\elt; e.g.
Osterbrock 1989) and the possibility that such abundances might be largely in
error, i.e. underestimated, in the case of strong fluctuations in temperature
(e.g. Peimbert 1967) or density (e.g. Rubin 1989) within the nebular volumes.
\emph{HST} observations of the optical \foiii\ nebular to auroral line ratio in
the Orion nebula, where the ADF is small, revealed residual small scale
temperature fluctuations (O'Dell et al. 2003), whose origin has not been fully
elucidated. However, it has now become clear that invoking in all generality
the presence of such fluctuations in a chemically homogeneous medium fails to
account consistently for the observed behaviour of the ADFs; chiefly the fact
that these do not correlate with the excitation energies or critical densities
of the lines in question. This is true for PNe (Liu et al. 2000; Tsamis et al.
2003b, 2004) and for \hii\ regions (Tsamis et al. 2003a, hereafter T03). In
parallel, the recent downward revision of the solar oxygen abundance (Allende
Prieto et al. 2001) has removed the long-lasting impression that the oxygen
abundances of galactic PNe were lower than the Sun's and possibly therefore
only lower limits: mean PN oxygen abundances derived from forbidden lines
(Kingsburgh and Barlow 1994; Exter et al. 2004) are now in good agreement with
solar values. Finally, recent spectroscopic results making use of the weakly
\elt-dependent ratios of \oii\ ORLs have provided evidence for the presence of
a very low temperature ($\lesssim$\,10$^3$\,K) gaseous component embedded
within both typical PNe (Tsamis et al. 2004) and atypical (`born-again') PNe
(Wesson et al. 2003).

The question of whether the supersonic velocities and induced shocks from
Wolf-Rayet stellar winds and SNRs (e.g. Chu \& Kennicutt 1994) might produce
significant temperature fluctuations within giant \hii\ regions (GHIIRs),
unaccounted for by static photoionization codes, warrants further analysis:
however, as long as the ORL/CEL abundance discrepancies are found to be
uncorrelated with the excitation energies of the various collisionally excited
lines, the answer to the above question does not bear heavily on the ORL/CEL
abundance discrepancy issue.

The currently most viable explanation for both the ORL/CEL abundance dichotomy
and the very low temperatures derived from recombination lines or continua,
posits the existence within nebulae of a gas component enhanced in heavy
elements and therefore efficiently cooled by emission in fine structure CELs
(Liu et al. 2000; P\'{e}quignot et al. 2002; Tsamis et al. 2004). This material
which is traced by the emission of weak heavy element ORLs, is intermingled
with a more normal composition, hotter ($\sim$\,10$^4$\,K) medium, traced by
the emission of strong CELs. The above picture has gained in credibility since
it has been supported by dual-abundance photoionization models of PNe
(P\'{e}quignot et al. 2002, 2003; Tylenda 2003; Ercolano et al. 2003; Tsamis \&
P\'{e}quignot 2005 and to be submitted).

From the standpoint of chemical abundance analyses, an important result of the
dual-abundance model is that the high metallicity component comprises only a
small fraction ($\sim$\,1 per cent) of the total ionized nebular mass in the
objects considered so far, so that the heavy-element content of the nebulae is
closer to the one given by the classical forbidden-line analysis (P\'{e}quignot
et al. 2002; Tsamis and P\'{e}quignot 2004). This contrasts with the classical
temperature fluctuation paradigm, which invariably leads to CNONe abundances
closer to those given by the ORL method for PNe (Ruiz et al. 2003; Peimbert et
al. 2004) and for \hii\ regions (Peimbert 2003, hereafter P03; Esteban et al.
2005 and references therein). Since emission-line studies of \hii\ regions
provide a wealth of information regarding elemental abundances in extragalactic
systems and therefore have a considerable bearing on cosmic chemical evolution
studies, the question of which description provides the right answer is a
pressing one.

The co-existence, in PNe, of gas phases with very different compositions and
temperatures may plausibly result from the ejection of material from the
evolving central star whose surface composition changes as nuclear reactions
and mass loss proceed, as in e.g. the `born-again PN' scenario (Iben et al.
1983). This is perhaps more unexpected in \hii\ regions which are recently
ionized zones of the `normal' interstellar medium (ISM). These objects,
however, are not born \emph{ex nihilo}. Their chemical composition depends on
the past history of the host galaxy: the thorough mixing of the ISM is a long
process in which successive generations of massive stars may be involved (e.g.
Tenorio-Tagle 1996). They are also dynamic systems, hosting potential sources
of chemical inhomogeneities, such as supernova remnants (SNRs), evolved massive
O/WR-type stars subject to extensive mass-loss via stellar winds, and
photo-evaporating protoplanetary disks.\footnote{30\,Dor contains both SNRs
(such as N157B; e.g. Lazendic et al. 2003) and numerous Wolf-Rayet stars (see
Parker 1993; Crowther \& Dessart 1998), while a plethora of `proplyds' have
been identified in e.g. the Orion nebula.}  In the face of recent studies which
have uncovered fairly large ADFs ($\sim$2--5 for \opp) for galactic and
extragalactic \hii\ regions (Esteban et al. 2002; T03), efforts should be
directed at clarifying the situation.

In this paper we present a photoionization model study of the 30 Doradus nebula
for which high quality, multi-waveband spectroscopy has become available in
recent years, including the heavy element ORLs of interest. We would like to
investigate whether, by analogy with PNe, the existence of chemical
inhomogeneities could provide a solution to the ORL/CEL problem for this class
of nebulae as well, and thus help us choose between the two conflicting
abundance sets. In Section~2 we collate available spectroscopic data pertinent
to this study, while in Section~3 we present the models and results. Finally,
in Section~4 we discuss the implications of this work for emission-line studies
of photoionized nebulae and consider various hypotheses on the likely nature
and origin of the postulated chemical inhomogeneities in \hii\ regions.

\section{Observations}

At an adopted distance of 50\,kpc, the 30 Doradus nebula in the LMC is the
closest extragalactic giant \hii\ region in our cosmic neighbourhood. Its full
extent in \ha\ emission is well over a degree in diameter ($\simeq$1\,kpc). It
is photoionized by a tremendous amount of Lyman (Ly) continuum radiation whose
total flux has been estimated to be up to $\sim$10$^{52}$ photons s$^{-1}$ from
radio observations of the entire nebula (Mills, Turtle \& Watkinson 1978),
while almost half of this (4.2$\times$10$^{51}$\,photons s$^{-1}$) arises from
within the inner 10\,pc, provided by 117 massive OB and WR-type stars of the
R136 cluster in the nebula's core (Crowther \& Dessart 1998). The integrated
(reddened) \ha\ flux of the central 30$'$ is 1.71$\times$10$^{-8}$
erg\,cm$^{-2}$\,s$^{-1}$ (Kennicutt \& Hodge 1986), corresponding to
4.2$\times$10$^{51}$\,photons s$^{-1}$ in the Ly continuum.


Over the years the nebula has been observed extensively from UV to radio
wavelengths. In this study we made use of optical emission-line fluxes by
P03 (UVES/VLT 8.2-m) and T03 (EMMI/NTT 3.6-m); UV fluxes by Dufour
et al. (1982; \emph{IUE}) and Garnett et al. (1995; \emph{HST} FOS); IR fluxes
by Vermeij et al. (2002; \emph{ISO} SWS \& LWS), except for the \farii\
6.98\,$\mu$m line noted by Giveon et al. (2002); and radio fluxes by Filipovic
et al. (1995). A compilation of these is provided in Table~3 [col. 3;
dereddened intensities expressed in units such that $I$(\hb)$=$10$^3$].

The \emph{ISO} SWS and LWS data are for the position `30\,Dor\,\#3' of Vermeij
et al. (2002); the positions observed optically by T03 and P03 fell
within this pointing and the 80$''$ LWS circular beam. Peimbert's
3$''$$\times$10$''$ echelle slit was placed on a bright filamentary `ridge'
located 64$''$ north and 60$''$ east from HD~38268 ($=$ R136) and was contained
within the largest SWS aperture, while the long-slit of T03 fell about
20--30$''$ to the west, missing the 20$''$$\times$33$''$ SWS aperture and
passing through the multiple stars R139 and R140\footnote{Both R139 and R140
contain WR components; see their spectrograms by T03.} on an approximate NE-SW
direction. From a comparison of the intensities of 8 \hi\ lines (Br$\alpha$,
Br$\beta$ and Pf$\alpha$--$\zeta$) listed by Vermeij et al. with the
theoretical recombination-line ratios by Storey \& Hummer (1995), the estimated
\hb\ flux, corrected for interstellar extinction, falling into the
14$''$$\times$20$''$ SWS aperture is
1.27$\times$10$^{-10}$\,erg\,cm$^{-2}$\,s$^{-1}$. The line fluxes between
12--180\,$\mu$m from Vermeij et al. were scaled down by factors of 1.35 (SWS,
12--27.5\,$\mu$m), 2.35 (SWS, 27.5--45.2\,$\mu$m) and 7.35 (LWS,
45--180\,$\mu$m), based on the relative sizes of the four rectangular SWS
apertures and the need for flux continuity in the spectral continuum, and then
cast in units relative to \hb. Regarding UV data, the dereddened intensity of
\ciii] $\lambda$1909 is from observations in the years 1979--1980 (Dufour et
al. 1982), and was adopted instead of more recent \emph{HST} observations by
Garnett et al. (1995a) taken through the 1$''$ FOS circular aperture. The
10$''$$\times$20$''$ elliptical aperture of the \emph{IUE} was centered on a
position roughly identical to that targeted by the SWS. We used the measured
line strengths, relative to $\lambda$1909, by Garnett et al. (1995a) for the
\oiii] $\lambda$1666 and \niii] $\lambda$1750 intensities (which are therefore
tentative only), while the Si~{\sc iii}] $\lambda$$\lambda$1883, 1892 fluxes
were from the $F$(Si~{\sc iii}])/$F$(C~{\sc iii}]) ratio measured by Garnett et
al. (1995b) with the FOS.

The heavy element ORLs utilized in this work include the \cii\ $\lambda$4267
line and the multiplets \nii\ $\lambda$5679 (V3), \oi\ $\lambda$7773 (V1),
\oii\ $\lambda$4075 (V10), \oii\ $\lambda$4340 (V2) and \oii\ $\lambda$4650
(V1) (co-added intensities; standard theoretical relative line ratios were used
to estimate unobserved multiplet components where necessary; see e.g. Liu et
al. 1995, 2000; Tsamis et al. 2004). Upper limits were adopted for the
undetected \nii\ $\lambda$4041, \neii\ $\lambda$4392 and \mgii\ $\lambda$4481
lines.

For this study the adopted total dereddened \hb\ intensity from 30 Dor is
1.19$\times$10$^{-8}$\,erg\,cm$^{-2}$\,s$^{-1}$, in line with the
aforementioned \ha\ estimate by Kennicutt and Hodge (1986) for the central
30$'$; the adopted logarithmic reddening constant was $c$(\hb) $=$ 0.44. T03
and Vermeij et al. (2002) found values of 0.41 and 0.48 from fixed long-slit
and drift-scan slit spectra respectively. These estimates compare favourably
with a mean extinction of about 0.5\,dex derived from \ha/\hb\ ratio maps of
the central 6$'$$\times$6$'$ region by Lazendic et al. (2003).
The optical line intensities listed in Table 3 are essentially taken from
P03, who dereddened the fluxes using $c$(\hb) $=$ 0.92 (the maps of
Lazendic et al. do indicate higher than average extinction at the position
observed by P03). The NTT long-slit spectra of T03 which sampled a more
extended `slice' of the nebula were consulted for consistency checks. The UV,
IR and radio fluxes, although carefully calibrated relative to \hb, do not
necessarily sample exactly the same emission regions targeted by the optical
observations.

\section{Photoionization modelling}

\subsection{The code}

The modelling is undertaken using the detailed photoionization code {\sc nebu}
(e.g. P\'{e}quignot et al. 2001) in which the radiative transfer is computed
using the outward only approximation along 20 directions in spherical symmetry,
and the ionization state and plasma temperature are determined by solving the
relevant equilibrium equations. A model is built by combining a user-defined
number of sectors extracted from different spherically symmetric models sharing
the same central source of primary ionizing radiation, each with its own
covering factor $\omega/4\pi$ of the source, radial hydrogen density
distribution $N_{\rm H}(r)$, gas filling factor $\epsilon$, and set of
elemental abundances relative to H.
The code version used here does not include an interstellar dust component.



\subsection{Modelling strategy}

\subsubsection{The `average filament'}

30\,Dor is much larger than R136 and can be viewed as a gaseous shell
photoionized by a point-like source located at the centre of
an inner cavity. Published images reveal the complex, frothy
appearance of the nebula, whose overlapping loops, arcs, and shells are
believed to have originated in interacting stellar-wind outflows and expanding
SNRs (e.g. see the 3- and 6-cm radio and \ha, \hb\ images by Lazendic, Dickel,
\& Jones 2003 and references to imaging and kinematical studies therein). The
mean square root electron density $\langle$\eld$^2$$\rangle$$^{\frac{1}{2}}$ of
30\,Dor, derived from the \hb\ flux of the central 30$'$ under the assumption
of a uniformly filled sphere, is
two orders of magnitude smaller than the \eld\ derived from the \foii\ and
\fsii\ doublet ratios. This indicates that the bulk of the emission arises from
a small fraction of the total volume, as suggested also by the highly
filamentary appearance of the nebula. This common feature of GHIIRs is often
handled with the concept of a `filling factor', $\epsilon$. In the classical
form of this concept, the emitting gas is assumed to belong to
\emph{infinitesimal}, optically thin clumps distributed more or less uniformly
between an inner and an outer radius, and occupying altogether a fraction
$\epsilon$ of the nebular volume (e.g. Osterbrock 1989). In this approximation,
which aims at emphasizing the observed geometrical extent of the \hii\ region,
the high excitation gas generally lies close to the energy source whereas the
lower excitation gas is found farther out. This is usually not borne out by
observation: although shallow radial gradients may exist throughout the nebula,
filaments tend to emit both high and low ionization lines at whatever distance
from the ionizing source.\footnote{This is indicated from an analysis along
$\sim$150$''$ of the slit of T03 which was placed roughly along the nebula's
radial direction and crossed several filamentary nebulosities (cf. fig.\,~5 of
that paper).} The very existence of a filamentary structure contradicts the
classical concept. As confirmed by photoionization modelling, given the primary
ionizing flux and the gas density of 30\,Dor, matter becomes optically thick to
ionizing radiation over distances which are much smaller than the nebula and
similar to the typical thickness of the observed filaments (e.g., compare the
spatial thickness of the model nebula with its radius in Table~2). Large-scale
radial gradients, if present, are likely due to a global variation of the
ionizing parameter with position, e.g. in the case that the average density of
the different filaments does not scale as the inverse square of their distance
to the ionizing source.

As an alternative to the classical description, we consider \emph{macroscopic},
essentially radiation-bounded filaments which are distributed throughout the
nebula. Each one of them produces a radially extended shadow, which emits much
less than the filament itself since it is only subjected to the weak, extremely
soft, diffuse field from other filaments. Here, this alternative description is
adopted in its simplest first approximation, namely in terms of one
representative `average filament'. Given its small geometrical thickness, the
radial structure of the filament can be treated assuming spherical symmetry
without loss of generality. As a matter of fact, the adopted geometry enters
only in the computation of the diffuse field, and there only to a limited
extent since the diffuse field is either relatively unimportant in the limiting
case of small opacity/optical depth, or very local (i.e. fully independent of
geometry) in the opposite limiting case. Depending on the overall covering
factor of the entire set of filaments, ionizing radiation may or may not leak
out from the nebula. In this work, it is assumed that the overall covering
factor is unity, and the power of the ionizing source is the one inferred from
the integrated \ha\ emission over 30$'$, i.e. a large fraction of the total
emission. This amounts to computing the emission of a fully spherical,
geometrically thin shell with no escape of the primary and diffuse field.
Adopting the spectrum secured by P03 as representative of the whole nebula
simply means that the properties of our `average filament' will be somehow
weighted by the regions effectively observed.

\setcounter{table}{0}
\begin{table}
\begin{center}
\caption{Model parameters and observational constraints.}
\begin{tabular}{ll}
\noalign{\vskip3pt} \noalign{\hrule} \noalign{\vskip3pt}
Parameter     &   Constraint (flux or flux ratio)                                                     \\
\noalign{\vskip3pt} \noalign{\hrule} \noalign{\vskip3pt}
$c$(\hb)      & \ha/\hb\ (taken from P03)                                                                      \\
$L_\star$     &  absolute flux $I$(\hb)                                       \\
$T_{\rm eff}$&\foiii\ $\lambda$4959 + $\lambda$ 5007                                                 \\
$\delta$\,(4\,ryd)$^a$ &\heii\ $\lambda$4686  (for blackbody models only)                                      \\
\noalign{\vskip2pt}
$R_{\rm in}$  &\foiii\ ($\lambda$4959 + $\lambda$5007)/\foii\ ($\lambda$3726 + $\lambda$3729)  \\
$P_{\rm in}$        &\foiii\ 52\,$\mu$m/88\,$\mu$m  or \fariv/\fariii                     \\
$P_{\rm out}$ &\foii\ $\lambda$3726/$\lambda$3729                                                    \\
\noalign{\vskip2pt}
He            &mean of \hei\ $\lambda\lambda$4471, 5876                                 \\
C             &\fciii\ $\lambda$1908                                                                   \\
N             &\fnii\ $\lambda$6548 + $\lambda$6584                                                    \\
O             &\foiii\ $\lambda$4363/($\lambda$4959 + $\lambda$5007)                              \\
Ne            &\fneiii\ $\lambda$3869 + $\lambda$3968                                                           \\
Mg            &mean of Mg~{\sc i}] $\lambda\lambda$4562, 4571                                           \\
Si            &mean of Si~{\sc iii}] $\lambda\lambda$1883, 1892                                             \\
S             &mean of \fsiii\ $\lambda$$\lambda$6312, 9069, 9532                                           \\
Cl            &mean of \fcliii\ $\lambda\lambda$5517, 5537                                                      \\
Ar            &\fariii\ $\lambda$7135 + $\lambda$7751                                                       \\
Fe            &\ffeiii\ $\lambda$4658                                                                           \\

\noalign{\vskip3pt} \noalign{\hrule} \noalign{\vskip3pt}
\end{tabular}
\begin{description}
\item[$^a$] Scaling factor of the blackbody flux at $\geq$4\,ryd (see text).
\end{description}
\end{center}
\end{table}

\subsubsection{Model parameters}

Computations are therefore done assuming a smooth small-scale density
distribution, i.e. a gas filling factor of unity. The pair of (\elt, \eld)
varies according to the following general law for a variable gas pressure $P$,
given here as a function of the radial optical depth, $\tau$, at 13.6\,eV,

\begin{equation}
P(\tau)=\frac{P_{\rm in}~+~P_{\rm out}}{2}~+~\frac{P_{\rm out}~-~P_{\rm
in}}{\pi}~\tan^{-1}\Bigl[\kappa~\log \Bigl(\frac{\tau}{\tau_{\rm
\kappa}}\Bigl)\Bigr].
\end{equation}

\noindent The pressure is related to the gas temperature and density via the
ideal gas law. At the first step of the computation ($\tau = 0$) the initial
pressure is $P_{\rm in}$, while at the last step ($\tau = \infty$; in practice
$\gg$1) the final pressure is $P_{\rm out}$.  Eq.\,~1 introduces 3 free
parameters in the modelling process: $P_{\rm in}$, $P_{\rm out}$, and the
optical depth at which the transition from inner to outer pressures occurs
($\tau_{\rm \kappa}$). The `slope' of this transition is controlled by
$\kappa$, which is arbitrarily taken as $\kappa$ $=$ 10, ensuring a smooth
transition in all computations presented here. In practice, $\tau_{\rm \kappa}$
will have almost the same value 3.4~$\pm$~0.1 in all computations and the gas
distribution is practically controlled by two free parameters, $P_{\rm in}$ and
$P_{\rm out}$, along with the inner nebular radius. The realistic picture of a
filament core surrounded by a dilute halo or envelope dictates that $P_{\rm
in}$ $<$ $P_{\rm out}$. The adopted pressure law proved to be a convenient tool
for the exploration of the effects of the density distribution on the model
constraints and predictions; e.g. $P_{\rm in}$ could be constrained by the
\foiii\ 52-$\mu$m/88-$\mu$m line ratio and/or the \foiii
$\lambda$5007/52-$\mu$m ratio, or else some ionization-balance consideration
(e.g. \fariv/\fariii), while $P_{\rm out}$ was constrained by the \foii\
$\lambda$3726/$\lambda$3729 ratio.

Either a black body (BB) or a model stellar atmosphere spectrum were used to
simulate the spectral energy distribution (SED) of the primary ionizing source.
The \emph{CoStar} model stellar atmospheres (Schaerer \& de Koter 1997), which
include non-LTE treatment, line blanketing and stellar wind effects, were
adopted. The central source was controlled by a luminosity $L$ and a
temperature \teff. For the BB models, a scaling factor $\delta$ was applied to
the flux at $\geq$4\,ryd in order to fit \heii\ $\lambda$4686 (from T03). Given
the very small amount of radiative energy involved here, $\delta$ is of
negligible consequence for any other output or prediction of the model
($\delta$~$\simeq$~0.10 in adopted models). For the \emph{CoStar} models,
additional parameters would be (i) the metallicity of the stellar atmosphere,
taken to be the metallicity of the LMC, and (ii) the luminosity distribution of
stars in the central cluster. Since our aim was not to realistically model any
particular stellar cluster but only to evaluate the consequences of varying the
primary SED within reasonable limits (i.e. BB vs. \emph{CoStar}), stars with
luminosity of 10$^{39}$\,erg\,s$^{-1}$ were systematically adopted here, with
their number defining the total luminosity of the source.

In summary, there are 17--18 basic line intensities and intensity ratios which
constrain an equal number of model parameters: i.e. the reddening correction; 2
or 3 central source parameters ($L$, \teff, $\delta$); 3 nebular parameters
($R_{\rm in}$, $P_{\rm in}$, $P_{\rm out}$); and 11 elemental abundances
relative to H -- see Table~1. All other UV, optical and IR observables
(Table~3) were treated as `model predictions', whose relevance was evaluated in
terms of the employed atomic data, the observational (and calibration)
uncertainties, as well as more general astrophysical considerations.

\subsection{Chemically homogeneous models}

We initially constructed single-component, chemically homogeneous models.
Adopted parameters and derived physical conditions for two of these -- referred
to as S1 and S2 (with a BB and a \emph{CoStar} ionizing spectrum respectively)
-- are listed in Table~2. Chemically inhomogeneous, dual-component models D1
and D2 are introduced in Sect.~3.4. The models are radiation bounded (optical
depths at the end of each computation are given in Table 2). \emph{CoStar}
models have a larger inner radius and a smaller spatial thickness, driven by
the need to lower the ionization state of the gas which was otherwise
overestimated, as judged from the ionization balance of oxygen mainly, due to
the increased hardness of the primary spectrum. The density profile of model S1
is quite similar to the profile of the normal abundance component of model D1
(Sect.~3.4) plotted in Fig.\,~1. The ratios of predicted over observed line
intensities (hereafter the `departure ratios') are listed in Table 3 (cols. 4
and 5) and the overall quality of the fit (i.e. closeness to unity) can be
judged from those.

\subsubsection{Comments}

Models S1 and S2 provide us with a view of the influence of the primary
spectrum on the predictions. To first order these models were successful in
that it was possible to generate a set of parameters fulfilling all basic
requirements of Table~2, while many predictions compare well with observation
(Table~3). Many of the unsatisfactory departure ratios correspond to poorly
determined line intensities or upper limits. This overall agreement suggests
that basic assumptions are quite acceptable, while the atomic data and the
observations used are generally quite accurate. A commentary on remaining
discrepancies is of interest in order to evaluate their significance and
determine the \emph{astrophysical} relevance of the chemically homogeneous
description. Also, many features of S1--2 are common with the
following, more elaborate, models. Optical lines are considered first. \\

\noindent {\bf \hei.}\,~The $\lambda$7065 line appears overestimated, but can
be affected by telluric absorption: this line is 1.5 times stronger in the
spectrum of T03. Singlet \hei\ lines were computed assuming case~B
recombination (e.g. Osterbrock 1989). Except for the very weak $\lambda$4437
line, all of those singlet lines which are much weaker under case~A (they are
noted with a `B' in col. 3 of Table~3) are systematically overestimated by the
model, indicating that there is some departure from case~B in 30\,Dor. This is
possible even in optically thick nebulae since the resonance lines of \hei\ can
be absorbed by hydrogen atoms before being converted into optical \hei\ lines,
and this misfit has no bearing on either the quality of the observations or the
relevance of the model premises. The \hei\ $\lambda$4121 misfit is possibly due
to the observed line intensity being affected by blending with \oii\ V20
multiplet lines other than $\lambda$4119.22 listed by P03. The perfect fit to
$\lambda$3888 was obtained partly thanks to a secondary parameter, namely an
average `He$^+$ turbulent temperature', which is introduced to modulate
somewhat the optical depth in the 2$^3$S -- n$^3$P transitions, and is
proportional to the column density of the 2$^3$S metastable level (concerning
the \hei\ triplet transfer, see, e.g. Osterbrock 1989). The validity of this
description and the quality of the observations can be appreciated from (i) the
good fit to $\lambda$3187, also sensitive to optical depth, and (ii) the fact
that the adopted \hep\ turbulent temperatures (6, 3, 4, and
5 $\times$ 10$^4$\,K for S1, S2, D1, and D2 respectively) are very reasonable.\\

\noindent {\bf Low-ionization lines.}\,~Inspection of Table~3 shows that
emission from low-ionization species, such as \foi\ $\lambda\lambda$6300, 6363,
is rather well fitted even though model S2 overestimates the \foi\ flux by
$\sim$0.3\,dex. An overestimation of \foi\ may suggest that, contrary to our
assumption, the material is not strictly radiation-bounded along all radial
directions. It may also indicate that too many `hard' photons arise from the
source and are eventually absorbed at the ionization front (see also the
commentary on \fariv\ below). On the other hand, emission of the \fni\
$\lambda$5200 multiplet and the \fsii\ lines is underestimated by 0.7 and
0.3\,dex respectively in model S1. \emph{CoStar} model S2 reproduces the \fsii\
flux much better, without improving the fit to the \fni\ lines. Note the
overall agreement of the \fsii\ relative line intensities. The intensity of the
far-red \fsii\ multiplet was based on the strongest line \fsii\ $\lambda$10320,
re-calibrated by means of the 3 strongest neighbouring \hi\ and \hei\ lines,
which consistently point to an upward correction of the intensities by a factor
of 1.28 in this wavelength range of the VLT spectrum (in this range, lines with
dereddened intensities less than 0.1 in the present units lead to inconsistent
results). The collision strengths for \fni\ are questionable as recent results
by Tayal (2000) inexplicably differ from previous ones by as much as one order
of magnitude. Here, the values obtained by Berrington \& Burke (1981) were
used. A small wavelength interval including \fni\ $\lambda$5200 was
unfortunately not observed by T03.

Finally, the \fci\ $\lambda$9850 line is to a large extent excited by
recombination ($\sim$ 50 per cent on average for the whole nebula). The fact
that it is underpredicted is to be considered in connection with the poor fit
to the heavy-element ORLs (see below), even though the line can also arise from
the photodissociation region (PDR), which emits
lines like the strong \foi\ 63-$\mu$m and is not modelled here.\\

\noindent {\bf High-ionization lines.}\,~All models, S2 especially,
overestimate the emission of \fariv\ and probably \fcliv\ also (the latter line
is weak). The recombination coefficients for these species are poorly known,
and the adopted approximate empirical values were derived from an unpublished
model of the PN NGC\,7027 by P\'equignot. For example, the empirical
coefficient is 8 times the radiative coefficient in the case of Ar$^{3+}$
recombination. The mean fractional concentrations of Ar$^{+3}$ and Cl$^{+3}$
are moderate ($\sim$8 and 13\,per cent in model S1; $\sim$11 and 21 per cent in
model S2), and very sensitive to density as these are the highest ions present
in the gas. While line fluxes by P03 and T03 are generally in fair agreement,
the \fariv\ lines are striking exceptions, with the fluxes of T03 about 70\,per
cent larger, suggesting that (i) the concentration of Ar$^{+3}$ indeed varies
greatly with position in the nebula and consequently, (ii) an overestimation of
these line fluxes not exceeding $\sim$0.3\,dex in a model is probably not
physically significant. From trials it was found that BB models at constant
densities (\eld~$=$~300\,\cmt) overestimate the \fariv\ and \fcliv\ fluxes by
$\sim$50\,per cent only, whereas \emph{CoStar} models with a positive radial
density gradient overestimate this flux by up to 0.60\,dex (model S2). The
latter discrepancy would tend to invalidate the prominent continuum flux excess
(relative to a BB of the same \teff) shown for $h\nu$ $>$ 40\,eV by the
{\emph{CoStar} stellar atmospheres, in the way they were used here. Also, the
extended, low density halo of the high-excitation zone of our model filament
was tailored to fit the \foiii\ FIR ratio (models S1--2). The 80$''$ LWS lobe
however sampled a large area in which the mean density may be predictably lower
than the one called for by the \fariv\ UVES ratio; UVES sampled a small bright
filamentary region whereas LWS targeted an area of lower average surface
brightness (see fig.\,~2 of Vermeij et al. 2002). These discrepancies are
therefore understood in terms of conflicting density sampling of the IR and
optical density-sensitive ratios due to the different spatial coverage of each
spectral domain, and of the uncertainty regarding the exact shape of the
primary continuum at high energies (see also model D2 in
Section~3.4).\\

\noindent {\bf Intermediate-ionization lines.}\,~The \fsiii\ auroral
$\lambda$6312 and nebular $\lambda$9069 + $\lambda$9532 lines are imperfectly
matched. The relative intensity ratio of the nebular lines agrees with theory
within 8 per cent (Mendoza 1983), although these lines are often affected by
telluric absorption; considering this agreement and their large strength, more
weight was given to them than to $\lambda$6312 in the determination of S/H.
Calculations were performed here using the collision strengths by Galavis et
al. (1995) for \fsiii, \fariii\ and \fcliv, who devoted particular attention to
the fine-structure lines and then argued that the \fariii\ results were not
controversial (Galavis et al. 1998). The collision strengths for \fsiii\ may
not be of ultimate accuracy however: in a trial calculation using the data
published by Tayal \& Gupta (1999), all 3 optical \fsiii\ lines were almost
exactly matched using the previously determined S/H abundance ratio.

The \fnii\,$\lambda$5755 line is some 10 per cent high in both S models,
pointing to a possible weakness of these chemically homogeneous models (see
Sect.~3.4). Here the presumably accurate collision strengths by Hudson \& Bell
(2005) were adopted.\\

\noindent {\bf CNO ORLs.}\,~The optical recombination lines of C,N and O are
systematically underpredicted. This is the \emph{raison d'\^etre} of the
present study (Sects.~3.3.2 \& 3.4).\\

\noindent {\bf IR lines.}\,~The computed fine-structure ratios for the \fneiii\
and \fariii\ transitions are not sensitive to model assumptions in the
conditions of 30~Dor, and the imperfect match of these ratios cannot be taken
as an inadequacy of the astrophysical assumptions themselves. Whatever the
cause, these discrepancies suggest that a mismatch of $\sim$15\,per cent for
other fine-structure line ratios might not be significant. The discrepancy is
larger for \fariii, but \fariii\ 21.8-$\mu$m is one of the weakest IR lines. It
is found from both S1 and S2 models that the predicted \fneiii\ 15.6- and
\foiii\ 51.8-$\mu$m line fluxes support a higher Ne/O ratio than their optical
counterparts do (if only by 15--18\,per cent). In this regard, these
homogeneous models fail to exactly fit the optical and IR Ne and O lines
simultaneously. In contrast, the S1--2 models differ significantly for \fniii\
57-$\mu$m, with the \emph{CoStar} computation leading to much more consistent
results. The collision strengths used for \fniii\ in these computations are
from Stafford et al. (1994). Were we to use the larger values obtained
previously by Blum \& Pradhan (1992), the difficulty met with \fniii\ in BB
models would be exacerbated. Nonetheless, the most obvious failure of these
models is the underprediction of \fneii. The IR lines are reconsidered
in Section~3.4.2.\\

\subsubsection{Two-sector models}

Chemically homogeneous models comprising two sectors were also investigated.
The idea (which has been tried in the past) was to check whether the different
mean densities of the two model sectors would induce significant global
temperature variations and whether this would result in a better fit for the
CNO ORLs. A simple approach was taken by using the same smooth scale density
law for both sectors, but imposing different densities (by a factor of 2) at
the innermost point of each. Vastly different starting values were not
supported by the observed density line ratio diagnostics. The model fitting was
as good as for model S1 with the same abundances for all elements. The fit to
the heavy element ORLs did not improve at all however (e.g. \cii\ $\lambda$4267
still underestimated by 0.3\,dex). The electron temperature difference between
the two components amounted to 120\,K while the mean density differed by a
factor $\sim$2. Resulting global temperature fluctuation parameters for
representative ions were \tsq(\hp) $=$ 0.0038 and \tsq(\opp) $=$ 0.0026.
Similar low \tsq\ factors were reported by Kingdon \& Ferland (1995) who
investigated chemically homogeneous models incorporating a much broader range
of density variations.

\subsubsection{Ionizing spectrum characteristics and model radii}

The Ly continuum ionizing flux of model S1 is 7.55$\times$10$^{51}$
photon\,s$^{-1}$, comparable to the Lyc flux of 4.2$\times$10$^{51}$
photon\,s$^{-1}$ derived by Crowther \& Dessart (1998) emanating from within a
10\,pc radius from the massive core cluster R136, or that of
$\sim$1.2$\times$10$^{52}$ photon\,s$^{-1}$ obtained by Israel \& Koornneef
(1979) for the inner 3$'$$\times$3$'$ (45\,pc$\times$45\,pc) of the nebula. By
construction, the total \hb\ flux computed by the model fits the measurement of
Kennicutt \& Hodge (1986) with a circular 30$'$ aperture (i.e. a physical
radius of $\sim$218\,pc). The absolute radio fluxes are also well accounted for
since the effective beam sizes used in the radio observations of Filipovic et
al. (1995) are effectively contained within the above 30$'$ aperture.

Regarding the shape of the ionizing spectrum, the black-body temperature of
49.3\,kK corresponding to model S1 falls within the \teff\ range of 43--51\,kK
derived by Martin-Hernandez et al. (2002) from the known, observed 30\,Dor
stellar spectral types using the \teff--spectral type calibrations of Vacca et
al. (1996) and Martins et al. (2002). The \teff\ obtained using the
\emph{CoStar} stellar atmosphere models however is 37--38\,kK only, again in
excellent agreement with the findings of Mart{\'{\i}}n~Hern{\'a}ndez et al.
(2002) who used those models and the {\sc cloudy} photoionization code to
derive the stellar properties of 30\,Dor from the same \emph{ISO} observations
that were used in this work.

The outer radii of these models span a range of 27--35\,pc (Table~2),
equivalent to 111--144$''$. This is in very good agreement with the actual
position of the UVES slit at a \emph{projected} distance of 88$''$ from R136.
Furthermore, the derived thickness of our `average filament' is 2--3\,pc
corresponding to 8--12$''$ at 50\,kpc. The filament of model D2 (Section~3.4)
is thinner being roughly 1\,pc wide (see Fig\,~1). Inspection of the \hb\ image
of the nebula presented by Lazendic et al. (2003), shows that the UVES slit
fell on a prominent filament with an apparent width of about 8$''$. These
findings suggest that our approach is successful inasmuch as, being based
exclusively on observed nebular line intensities, the models correctly
\emph{predict} both the global ionizing source properties and the approximate
spatial characteristics -- distance from R136 and extent -- of the region
sampled by spectroscopy.

The gas phase abundances of these chemically homogeneous models are discussed
later in Section~3.6 along with our preferred 30\,Dor gaseous abundances.
Taking into account the above remarks and the commentary of Sect.~3.3.1, we
find that the majority of lines, both CELs from heavy elements as well as \hi\
and \hei\ ORLs, can be reasonably well fitted, with the crucial exception of
the CNO ORLs: predictions for these were found to be identical across models,
i.e. consistently underestimating the observed ORL fluxes (Table 3). The
behaviour of these lines under widely different modelling assumptions such as
constant or variable gas density, and black body or stellar atmosphere ionizing
spectrum, necessitates a different approach to the modelling process. This is
introduced in the following section in the form of a dual abundance model
nebula comprising two sectors across which the elemental abundances (and
therefore other properties such as density and temperature also) vary.

\begin{figure}
\centering \epsfig{file=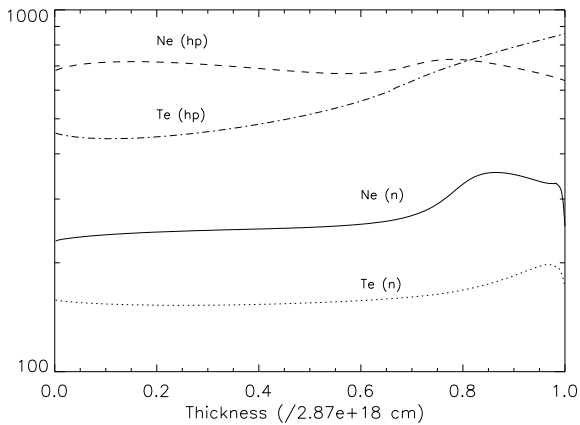, width=8.6 cm, clip=,} \caption{Electron
density distribution, \eld, and temperatures \elt\ (scaled down to 0.02\elt\
and 0.14\elt\ respectively for the normal (`n') and H-poor (`hp') phases) of
the dual abundance model D2.} 
\end{figure}

\setcounter{table}{1}
\begin{table*}
\centering
\begin{minipage}{150mm}
\caption{Model parameters and physical properties.}
\begin{tabular}{lcc@{\hspace{10mm}}cccccc}
\noalign{\vskip3pt} \noalign{\hrule} \noalign{\vskip3pt}
Model                                     &S1        &S2         &\multicolumn{3}{c}{D1}    &\multicolumn{3}{c}{D2} \\
                                          &          &           &H-poor  &Normal  &Mean       &H-poor  &Normal  &Mean      \\
\noalign{\vskip3pt} \noalign{\hrule}\noalign{\vskip3pt}
Primary spectrum                          &Planck    & $CoStar^a$ &\multicolumn{3}{c}{Planck}   &\multicolumn{3}{c}{$CoStar^a$}\\
Eff. temperature, \teff\ (kK)             & 49.3     & 37.6      &\multicolumn{3}{c}{51.6}      &\multicolumn{3}{c}{38.0}      \\
Luminosity, $L$ (10$^8$\,L$_{\odot}$)     & 1.12     & 1.98      &\multicolumn{3}{c}{1.08}      &\multicolumn{3}{c}{1.90}      \\
    \noalign{\vskip3pt}
Inner radius, $R_{\rm in}$ (10$^{19}$\,cm)& 7.80     &10.15      &\multicolumn{3}{c}{8.25}      &\multicolumn{3}{c}{10.50} \\
Thickness (10$^{18}$\,cm)                 & 8.39     & 5.07      &\multicolumn{3}{c}{8.23}      &\multicolumn{3}{c}{2.87}  \\
$P_{\rm in}$ (10$^{-10}$\,CGS)            & 4.16     & 4.16      &\multicolumn{3}{c}{4.00}      &\multicolumn{3}{c}{6.00}  \\
$P_{\rm out}$ (10$^{-10}$\,CGS)           &12.20     &12.40      &\multicolumn{3}{c}{11.1}      &\multicolumn{3}{c}{11.4}  \\
Covering factor, $\omega/4\pi$            & 1.00     & 1.00      &0.080    & 0.920  & 1.00$^b$  & 0.086    & 0.914  &  1.00$^b$ \\
Filling factor, $\epsilon$                & 1.00     & 1.00      &0.105    & 1.000  &   --      & 0.072    & 1.000  &  --  \\
Inner \hyd\ (\cmt)                        & 140      & 145       & 343     & 136 &  --          & 614    & 211     &  --   \\
$\langle$\eld$\rangle$ (\cmt)             & 201      & 204       & 440     & 192 &  198         & 696    & 271     &  279  \\
$\langle$\hyt$\rangle$ (K)                &9962      &9927       &5026     &9942 & 9716         &3987    & 9895    & 9654  \\
\tsq(\hp)                                 & 0.0045  & 0.0064     &0.0696   & 0.0054 & 0.0175    & 0.0544   & 0.0072 &0.0224  \\
\tsq(\opp)                                & 0.0031  & 0.0046     &0.0386   & 0.0039 & 0.0715    & 0.0338   & 0.0053 &0.0779  \\
$\tau$~(13.6\,eV)                         &  90      &  92       &  29     &  90 &  --          &   9    &    34     &  --   \\
\hb\ fraction                             & 1.00     & 1.00      &0.081    & 0.919  & 1.00$^b$  & 0.084  & 0.916    & 1.00$^b$ \\
Gaseous mass (10$^3$\,M$_{\odot}$)        & 151.     & 149.      &3.51     & 144.&  147.$^b$    &1.96    & 106.   &  108.$^b$ \\
   \noalign{\vskip2pt}
He/H                                      &0.0858    &0.0854     &0.1000  &0.0831  &0.0835     &0.1000  &0.0825 &0.0828  \\
   \noalign{\vskip2pt}
C/H \, \ \ $\times$10$^5$                 & 4.33     & 4.60      &66.0    &4.63    &5.98       &72.3    &4.96   &6.09   \\
N/H \, \ \ $\times$10$^5$                 & 1.74     & 1.42      &15.3    &1.74    &1.78       &18.8    &1.46   &1.75   \\
O/H \, \ \ $\times$10$^5$                 &22.90     &23.20      &200.0   &24.40   &28.30      &200.0   &25.40  &28.30  \\
Ne/H   \ $\times$10$^5$                   & 4.33     & 4.00      &44.0    &4.52    &5.39       &60.0    &4.38   &5.31   \\
   \noalign{\vskip2pt}
Mg/H  \ $\times$10$^7$                    & 52.0     & 39.5      &200.   &50.8    &54.7        &280.   &40.0   &44.3  \\
Si/H \,  \ $\times$10$^7$                 & 25.2     & 28.0      & 35.   &28.5    &28.6        &217.   &31.0   &34.0  \\
S/H \, \ \ $\times$10$^7$                 & 61.0     & 68.0      & 84.   &68.0    &68.4        &455.   &65.0   &71.4  \\
Cl/H   \ $\times$10$^7$                   & 0.59     & 0.69      & 0.8   &0.66    &0.66        & 5.0   &0.72   &0.79  \\
Ar/H  \ $\times$10$^7$                    & 12.0     & 12.6      &16.    &13.1    &13.2        & 92.   &13.2   &14.5  \\
Fe/H   \ $\times$10$^7$                   & 11.7     & 11.0      &17.    &13.7    &13.8        & 81.   &11.6   &12.7  \\
   \noalign{\vskip2pt}
$Y$                                       & 0.253    & 0.252     &0.274  &0.247   &0.248       &0.273  &0.246  &0.246  \\
$Z$                                       & 0.005    & 0.005     &0.036  &0.004   &0.005       &0.040  &0.004  &0.006  \\
   \noalign{\vskip3pt} \noalign{\hrule}\noalign{\vskip3pt}
\end{tabular}
\begin{description}
\item[$^a$] Stellar atmosphere model of 10$^{39}$\,erg\,s$^{-1}$ luminosity and typical LMC metallicity.
\item[$^b$] Sum of components instead of mean.
\end{description}
\end{minipage}
\end{table*}

\setcounter{table}{2}
\begin{table*}
\centering
\begin{minipage}{160mm}
\caption{Comparison of predictions from models S1, S2, D1, and D2. The
observed, dereddened intensities are in units such that $I$(\hb)$=$10$^3$,
except the radio continuum fluxes whose units are Jy; the model predictions are
given as the ratio of predicted over observed values in each case.}
\begin{tabular}{lrrcccc}
\noalign{\vskip3pt} \noalign{\hrule} \noalign{\vskip3pt}
Line    &$\lambda$ (\AA)$^a$ &$I_{\rm obs}$$^b$   &S1          &S2           &D1    & D2\\ 
\noalign{\vskip3pt} \noalign{\hrule}\noalign{\vskip3pt}                                 
\multicolumn{7}{c}{H, He recombination lines, optical and radio continua}\\

\hb     & 4861.3         &1000$^c$   &1.00         &1.00             &1.00         &1.00        \\
\ha     & 6562.8         &2850       &1.02         &1.02             &1.02         &1.02                       \\
\hg     & 4340.5         & 483       &0.97         &0.97             &0.97         &0.97                       \\
\noalign{\vskip2pt}
\hei$^d$ & 3187.7        & 27.4       &1.05         &1.06             &1.06         &1.03\\
\hei\,{\sc (b)}    & 3613.6 &  4.02      &1.28         &1.27         &1.24         &1.22\\
\hei    & 3888.6        & 51.7       &1.00         &1.00             &1.00         &1.00\\
\hei\,{\sc (b)}    & 3964.7 & 8.61       &1.23         &1.22         &1.19         &1.18\\
\hei    & 4026.2        & 21.5       &0.94         &0.94             &0.92         &0.92 \\
\hei    & 4120.8        &  2.60      &0.73         &0.73             &0.70         &0.69 \\
\hei    & 4387.9        &  5.50      &0.96         &0.96             &0.95         &0.94 \\
\hei\,{\sc (b)}    & 4437.6     &  0.565     &0.81         &0.81     &0.78         &0.77    \\
\hei    & 4471.5        & 43.1       &1.01         &1.01             &1.00         &1.00 \\
\hei    & 4713.2        &  4.81      &1.12         &1.11             &1.06         &1.05\\
\hei    & 4921.9        & 11.31      &1.01         &1.01             &1.00         &1.00\\
\hei\,{\sc (b)}    & 5015.7     & 22.04      &1.22         &1.22     &1.19         &1.17 \\
\hei    & 5875.7        &121.2       &1.00         &1.00             &1.00         &1.00\\
\hei    & 6678.2        & 33.85      &0.99         &0.99             &1.00         &1.00\\
\hei    & 7065.2        & 28.30      &1.76         &1.75             &1.62         &1.66\\
\hei\,{\sc (b)}    & 7281.3     &  4.52      &1.42         &1.41     &1.37         &1.35     \\
\heii   & 4686.0        &  0.171     &1.00         &1.54             &1.00         &1.88\\
\noalign{\vskip2pt}
BJ/\hb  & 3646          &4.53        &0.97         &0.98             &1.03         &1.06 \\
\noalign{\vskip2pt}
cont.    & 1.40\,GHz     &37.6        &1.05         &1.06             &1.03        &1.02              \\
cont.    & 2.45\,GHz     &37.0        &1.04         &1.04             &1.02        &1.00              \\
cont.    & 4.75\,GHz     &35.8        &1.02         &1.01             &0.99        &0.98              \\
cont.    & 4.85\,GHz     &36.0        &1.01         &1.00             &0.98        &0.97              \\
cont.    & 8.55\,GHz     &34.8        &0.98         &0.98             &0.96        &0.94              \\
\noalign{\vskip2pt}
\multicolumn{7}{c}{Heavy-element recombination lines}\\

[C~{\sc i}] &9850.3+    &0.276       &0.37         &0.34              &0.84        &0.66  \\
\cii    & 4267.2        &0.919       &0.43         &0.42             &1.00         &1.00 \\
\noalign{\vskip2pt}
\nii    &4038.0        &$<$0.1       &0.53         &0.39             &0.83         &0.86 \\
\nii    &5679.0+       &0.186        &0.71         &0.52             &1.00         &1.00 \\
\noalign{\vskip2pt}
\oi     &7773.0+       &0.124:       &0.58         &0.55             &1.58         &1.85 \\
\oii    &4075.0+       &3.05         &0.62         &0.63             &0.97         &0.97 \\
\oii    &4341.0+       &1.54         &0.69         &0.70             &1.05         &1.05 \\
\oii    &4651.0+       &3.70         &0.66         &0.68             &1.01         &1.01 \\
\noalign{\vskip2pt}
\neii   &4394.0        &$<$0.2       &0.44         &0.44             &0.74         &0.65 \\
\mgii   &4481.2        &$<$0.1       &0.50         &0.37             &0.61         &0.55 \\
\noalign{\vskip2pt}
\multicolumn{7}{c}{Collisionally excited lines (UV and optical)}\\

C~{\sc ii}]&2322+ $\rbrace$ & 70:    &0.67           &0.86            & 0.65       &0.79 \\
\foiii     &+2321  $\rbrace$ & --     & --            & --             & --         & --  \\
C~{\sc iii}]&1907+      &150        &1.00           &1.00            & 1.00       &1.00 \\
\civ    &1549           & 12:       &0.31           &0.83            & 0.37       &0.69 \\
\noalign{\vskip2pt}
\fni     &5197.9       & 1.29       &0.28           &0.27             &0.33        &0.16 \\
\fni     &5200.2       & 0.693      &0.56           &0.53             &0.64        &0.29 \\
\fnii   &5754.6       & 1.87        &1.08           &1.10             &0.88        &0.98 \\
\fnii   &6583.5+      & 152         &1.00           &1.00             &1.00        &1.00 \\
N~{\sc iii}] &1744 +   &$<$15       &0.76           &0.59             &0.61        &0.56 \\
\noalign{\vskip2pt}
\foi     &6300.3+     & 12.5        &1.05       &2.20                 &1.37        &1.04 \\
\foii    &3726.1      &586          &1.00       &1.00                 &1.00        &1.00 \\
\foii    &3728.8      &637          &1.00       &1.00                 &1.00        &1.00 \\
\foii    &7319.8+     & 37.3        &0.93       &0.96                 &0.89        &0.96 \\
O~{\sc iii}] &1663    &15:          &2.16       &2.19                 &2.18        &2.20 \\
\foiii  & 4363.2      & 32.1        &1.00       &1.00                 &1.00        &1.00 \\
\foiii  & 5006.8+     &6790         &1.00       &1.00                 &1.00        &1.00 \\
\noalign{\vskip2pt}
\fneiii &  3869.1+     &444         & 1.00           &1.00            &1.00        &1.00 \\
\noalign{\vskip2pt}
Mg~{\sc i}]&4562.6     &0.716       & 1.05           &1.06            &1.06        &1.05       \\
Mg~{\sc i}]&4571.1     &0.617       & 0.94           &0.94            &0.94        &0.95      \\
\noalign{\vskip2pt}
Si~{\sc iii}] &1882.7   &20.6        &1.04            &1.04            &1.04  &1.05       \\
Si~{\sc iii}] &1892.0   &14.4        &0.96            &0.96            &0.96  &0.97       \\
\end{tabular}
\end{minipage}
\end{table*}

\setcounter{table}{2}
\begin{table*}
\begin{minipage}{160mm}
\centering \caption{{\it --continued}}
\begin{tabular}{lcccccc}
\noalign{\vskip3pt} \noalign{\hrule} \noalign{\vskip3pt}
Line    &$\lambda$ (\AA) &$I_{\rm obs}$   &S1          &S2         &D1        &D2 \\
\noalign{\vskip3pt} \noalign{\hrule}\noalign{\vskip3pt}
\fsii  &  4068.6      &  8.71      &0.77             &1.37            &0.82  &1.03\\
\fsii  &  4076.3      &  2.95      &0.74             &1.31            &0.79  &0.98\\
\fsii  &  6716.4      & 71.2       &0.68             &1.19            &0.74  &0.90\\
\fsii  &  6730.8      & 65.6       &0.65             &1.13            &0.70  &0.87\\
\fsii  & 10320.+      &  6.6:      &0.90             &1.60            &0.96  &1.20\\
\fsiii &  6312.0      & 18.1       &0.79             &0.80            &0.78  &0.75\\
\fsiii &  9530.6+     &938         &1.08             &1.09            &1.08  &1.09\\
\noalign{\vskip2pt}
\fclii      &8578.7+       &0.76:      & 1.04            &1.58        &1.08  &1.64\\
\fcliii     &5517.7        &4.49       & 1.00            &1.00        &1.00  &0.99\\
\fcliii     &5537.7        &3.32       & 1.00            &1.00        &1.00  &1.01\\
\fcliii     &8433.7+       &0.250      & 0.99            &1.00        &0.99  &0.99\\
\fcliv      &8045.6+       &0.494      & 2.58            &4.79        &3.08  &3.92\\
\noalign{\vskip2pt}
\fariii &  5191.8      &  0.663        &1.38             &1.38        &1.36  &1.34\\
 \fariii &  7135.8+     &148            &1.00             &1.00        &1.00  &1.00\\
\fariv  &  4711.4      &  1.31         &2.55             &3.97        &2.99  &3.14\\
\fariv  &  4740.2      &  0.986        &2.52             &3.92        &2.95  &3.13\\
\noalign{\vskip2pt}
\ffeii    &5158.8        &0.306          &1.12             &1.78      &1.36  &1.39\\
\ffeii    &7155.2        &0.207          &0.78             &1.21      &0.93  &1.07\\
\ffeiii   &4008.3        &0.284          &0.81             &0.82      &0.79  &0.81\\
\ffeiii   &4658.2        &4.89           &1.00             &1.00      &1.00  &1.00\\
\ffeiii   &4701.5        &1.25           &0.98             &0.99      &0.99  &0.99\\
\ffeiii   &4733.9        &0.405          &0.88             &0.88      &0.89  &0.89\\
\ffeiii   &4881.1        &1.51           &0.74             &0.74      &0.73  &0.74\\
\ffeiii   &4924.7        &0.341          &0.66             &0.66      &0.67  &0.67\\
\ffeiii   &5270.5        &2.44           &1.01             &1.01      &1.02  &1.02\\
\ffeiv    &6734.4        &0.228           &1.72            &1.52      &1.87  &1.44\\
\noalign{\vskip2pt}
\multicolumn{7}{c}{Collisionally excited lines (IR)}\\

\fnii  &121.8\,$\mu$m  &$<$2.7      &0.86           &0.86             &1.20        &1.31 \\
\fniii &57.3\,$\mu$m   &104         &1.51           &1.10             &1.67        &1.36  \\
\noalign{\vskip2pt}
\foiii  &51.8\,$\mu$m &1330         &1.06       &1.08                 &1.40        &1.34 \\
\foiii  &88.3\,$\mu$m &1250         &1.07       &1.08                 &1.29        &1.07 \\
\noalign{\vskip2pt}
\fneii  &12.8\,$\mu$m  & 99.0       & 0.39           &0.18            &1.05        &0.79 \\
\fneiii &15.6\,$\mu$m  &650         & 0.91           &0.91            &1.22        &1.41 \\
\fneiii &36.0\,$\mu$m  & 65.0       & 0.81           &0.80            &1.07   &1.21      \\
\noalign{\vskip2pt}
\fsiii &18.7\,$\mu$m  &320         &1.05             &1.05            &1.13  &1.53\\
\fsiii &33.5\,$\mu$m  &550         &0.82             &0.81            &0.87  &1.04\\
\fsiv  &10.5\,$\mu$m  &600         &1.62             &1.93            &1.83  &1.59\\
\noalign{\vskip2pt}
\farii  & 6.98\,$\mu$m &  2.4:         &0.94             &1.43        &1.03  &1.83\\
\fariii & 8.99\,$\mu$m &115            &0.95             &0.95        &1.01  &1.27\\
\fariii &21.8\,$\mu$m  & 11.5          &0.65             &0.65        &0.69  &0.85\\

\noalign{\vskip3pt} \noalign{\hrule}\noalign{\vskip3pt}
\end{tabular}
\begin{description}
\item[$^a$] `+' indicates that predictions are for the total intensity of the multiplet in each case.
\item[$^b$] `:' indicates uncertain flux and `$<$' an upper limit flux estimate.
\item[$^c$] Fit to dereddened, total \hb\ flux $I$(\hb) = 1.19$\times$10$^{-8}$\,erg\,s$^{-1}$\,cm$^{-2}$.
\item[$^d$] `{\sc (b)}' indicates a prediction for case B recombination of some \hei\ singlet lines (see text).
\end{description}
\end{minipage}
\end{table*}

\subsection{Dual component chemically inhomogeneous models}

Models that comprise two components which are intimately intermixed are now
considered. It is assumed that clumps or filaments of relatively H-poor gas are
embedded in the main body of the nebula, which consists of gas of approximately
`normal' LMC composition. It is also assumed that the clumped phase is
coextensive with the H-rich component. This latter assumption is a simple guess
suggested by evidence from the spatial analysis of T03, which revealed a
roughly uniform surface brightness ratio of the \cii\ $\lambda$4267 and \oii\
ORLs to \hb\ across a 160 arcsec swathe of the inner 30\,Dor region.
Technically, the results of 2 appropriately defined spherically symmetric
computations were combined. In order for a dual abundance model to be obtained,
a small sector of the normal composition shell was extracted and substituted by
a sector of the H-poor shell to the effect that the combined emitted \hb\ flux
remained almost unaltered. The gas filling factor in the main shell was kept at
unity, whereas the small filling factor of the H-poor shell was adjusted so
that the geometrical thickness of both shells were identical. The assumption of
intimate mixing over a scale much smaller than the shell size led to the
complementary notion of \emph{equal pressures for both components at similar
optical depths}. Given the identical geometrical extent of the two sectors and
the identical amounts of radiation they absorbed and reprocessed, the total
radiation field experienced by both were approximately similar at any optical
depth, and the predicted emission from the dual-sector model was very close to
the one which would arise from a single-sector model with many small H-poor
clumps uniformly embedded in the normal abundance gas throughout the whole
shell. The pair of (\elt, \eld) values varied along the radial direction
throughout both components according to a $P$($\tau$) law similar to that
adopted for the single component models (Fig.\,~1).

We consider two representative models: D1 which is built under a BB SED and has
fully radiation-bounded components, and D2 which is built under a \emph{CoStar}
SED. D2 was tentatively fine-tuned to fix the \foi\ excess of S2, by
considering a large, but finite optical depth in the normal component
(Sect.~3.3.1). The `one-filament approximation' is obviously coarse and the
adopted model filament must be seen as a best compromise. Specifically, for
$P_{\rm in}$ $\lesssim$ $P_{\rm out}$/4, the \fariv\ lines become unacceptably
strong, even with the proviso of Section~3.3.1, while for $P_{\rm in}$
$\gtrsim$ $P_{\rm out}$, \foiii\ 88-$\mu$m and \fsiii\ 33-$\mu$m become too
weak compared to \foiii\ 52\,$\mu$m and \fsiii\ 18\,$\mu$m respectively (see
Section~3.4.2). Accordingly, $P_{\rm in}$ $\sim$ $P_{\rm out}$/2 was adopted in
model D2. Other models with $P_{\rm in}$ $\sim$ $P_{\rm out}$/3 gave more
weight to the IR \foiii\ ratio. Models D1--2 should be considered in comparison
to their respective, one-component counterparts, S1--2.

\subsubsection{Constraining the metallicities of the model components}

For an increasing metallicity of the clumped component, its electron
temperature decreases due to more efficient cooling from CELs while, because of
the equal pressures, the electron density proportionally increases. The higher
density of the H-poor gas favours \npp\ and \opp\ recombination which produces
\np\ and \op. H$^0$ and He$^0$ are also enhanced there assisting in the charge
exchange process with \opp\ and \npp\ which produces extra singly ionized
species (Fig.\,~3) and boosts the overall efficiency of O~{\sc i} relative to
\oii\ ORL emission. The observed intensity of the very weak O~{\sc i}
$\lambda$7773 ORL being rather uncertain, an underestimation of the line may
not be significant, although a substantial overestimation is. O~{\sc i}
$\lambda$7773 therefore provides an upper limit to the O/H abundance ratio for
the posited clumped phase. On the other hand, for an O/H ratio significantly
less than this upper limit, the H-poor phase heats up and the collisional
excitation of \fnii\ $\lambda\lambda$6548, 6584 increases selectively due to
the low excitation energy of the lines. These two occurrences can help us to
bracket the abundances of the posited H-deficient regions (see also
Section~3.4.2).

Regarding model D1, the Si--Fe abundances of the clumped component were kept
approximately equal \emph{by mass} to those of the normal component, while
those of CNONeMg were enhanced. The ratio of oxygen mass fraction (relative to
H) of the two components is $m_{\rm O}^{1}$/$m_{\rm O}^{2}$ $=$
[(O/H)$_1$/(O/H)$_2$]($\mu_{\rm H}^{1}$/$\mu_{\rm H}^{2}$)$^{-1}$, where O/H
denotes abundance ratio by number and $\mu_{\rm H}^{i}$ is the mass per gram of
hydrogen. For model D1, (O/H)$_1$ = 2$\times$10$^{-3}$ was adopted after
trials and $m_{\rm O}^{1}$/$m_{\rm O}^{2}$ $=$ 7.6 was found. Then the CNONeMg
abundances in the clumps were adjusted iteratively until the observed
intensities of the ORLs were matched and without violating the adopted upper
limits in the case of undetected lines. The resulting ratios for the C, N, Ne,
and Mg mass fractions are 13.1, 9.5, 9.0, and 4.1 (the latter arbitrary)
respectively. Thus the H-poor phase of D1 has mass fractions of $X$ $=$ 0.690,
$Y$ $=$ 0.274, and $Z$ $=$ 0.036.

For model D2 the abundances of Mg--Fe were adopted to be 7 times larger in the
H-poor component than in the `normal' one; this enrichment factor is close to
that found for oxygen. Evidence in favour of enhanced Si--Fe in the H-poor
component is presented in Section~3.4.2. The resulting mass fractions of the
H-poor phase in model D2 are $X$ $=$ 0.687, $Y$ $=$ 0.273, and $Z$ $=$ 0.040.

Considering a more extreme enrichment of the clumped phase in heavy elements, a
model where the O/H abundance ratio in the metal-rich component was 50 times
higher than in the normal phase of model D1 (with Mg--Ar abundances equal by
mass in both components), led to much overestimated O~{\sc i} $\lambda$7773
emission (by 0.85\,dex), but to a Balmer jump approximately equal to that
observed.  A variant of this with Mg--Ar abundances also enhanced by 50 times
relative to those in the normal phase, overestimated the Balmer jump by 23\,per
cent, and resulted in much overestimated total O~{\sc i} $\lambda$7773 and
[Ar~{\sc ii}] 6.98\,$\mu$m line fluxes -- by prohibitive factors of $\sim$10
and $\sim$6 respectively. The fraction of total \hb\ flux emitted from the
clumps was 4.6\,per cent, the clump \oppt\ dropped to 978\,K, and the mass of
the clumped phase was reduced to 254$M_{\odot}$. The aforementioned
over-predictions for the O~{\sc i} and [Ar~{\sc ii}] lines did not improve in
trials where the computation involved a density bounded (as opposed to a
radiation bounded) H-poor region. We conclude that models where the H-deficient
phase has an O abundance larger than about 10 times (by number) that of the
ambient 30\,Dor medium are ruled out.

\subsubsection{Model predictions}

\begin{figure*}
\centering \epsfig{file=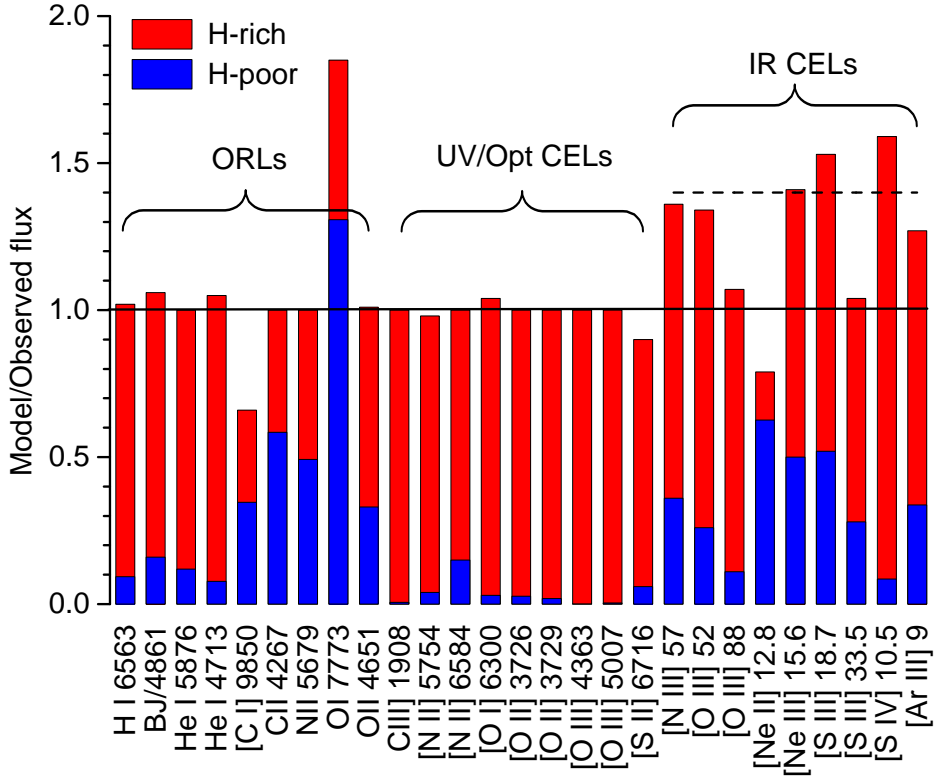, width=8.6 cm, clip=, angle=-90, scale=1.3}
\caption{The model over observed flux ratios (`departure ratios') for the
chemically inhomogeneous model D2 (Tables 2, 3): the relative contributions of
each nebular component to the individual line fluxes are colour-coded. Three
major groups of lines are identified. \fci\ $\lambda$9850 is emitted also from
the PDR which is not modelled here; \oi\ $\lambda$7773 is very weakly observed;
the deficit of \foiii\ 88- and \fsiii\ 33-$\mu$m is presumably due to the lines
being also selectively emitted by lower density background gas in the
\emph{ISO} fields of view. Amongst the IR lines shown, \fneii\ 12.8- and
\fsiii\ 33-$\mu$m are less accurately measured. The dashed line shows the
suggested re-calibration of the IR fluxes relative to the optical (see
Section~3.4.2 for details).}
\end{figure*}

The H-deficient phase of D1--2 is of lower ionization compared to the normal
phase: there is $\sim$2 times more H$^0$ (relative to \hp) and $\sim$4 times
more He$^0$ (relative to \hep) in the clump medium than in the diffuse medium
(Table 5). The clumped phase emits only about 8\,per cent of the total nebular
\hb\ flux, but it emits a more significant fraction of \fnii\ 6548, 6584-\AA,
\fniii\ 57\,$\mu$m, \foiii\ 52-, 88\,$\mu$m, \fneiii\ 15.6\,$\mu$m, and \foii\
3727-\AA; up to $\sim$30\,per cent for the optical and IR nitrogen lines and
somewhat less for the oxygen IR lines in model D1. In model D2 about
30--35\,per cent of the \fneiii\ 15.6-, \fsiii\ 18.7-, and \fariii\ 8.99-$\mu$m
fluxes are emitted from the cool H-poor component because of its enrichment in
Ne, S and Ar. The flux portion of the \cii\ $\lambda$4267 line, the \oii\
$\lambda$4650 V1 multiplet, and each of the \hei\ $\lambda\lambda$4471, 5876
and 6678 lines emitted from the clumped phase is $\sim$58, 32 and 10\,per cent
of the total respectively for both D1--2. The relative contributions of each
nebular component to the individual emission line fluxes are shown in Fig.\,~2
for a representative selection of ORLs, optical/UV CELs and IR CELs in model
D2.

Other than the CNO ORLs, inspection of Table 3 shows that the predictions for
\fci\ $\lambda$9850 and \fneii\ 12.8\,$\mu$m are also improved by about
0.3--0.4 and 0.4--0.6\,dex respectively in models D1--2 compared to the
chemically homogeneous models. The former line is efficiently excited largely
by radiative recombination\footnote{The collisional contribution to the line
excitation is $\sim$ 35 per cent in the H-poor component and $\sim$ 50 per cent
on average for the whole nebula.} of \cp\ in the clumped phase (see also
Section~3.3.1), while emission of the latter is favoured by the lower
temperature, as well as the 5 times higher concentration of \nep\ there (Table
5). The major flux portion of both these lines ($\sim$60--65\,per cent of the
total) arises from the H-deficient inclusions. The \hei\ line spectrum is even
better fitted in D1--2 models (concerning lines sensitive to departure from
case B and $\lambda$7065, see Sect.~3.3.1). Notably, due to its partial
emission from a cool medium, the predicted flux of the more sensitive to
collisional excitation \hei\ $\lambda$4713 line, is improved by 6 per cent.
Main features of the high excitation lines are as in S1--2 (Sect.~3.3.1), with
D1 somewhat worse than S1 and D2 somewhat better than S2. On the other hand, D2
is slightly less satisfactory than D1 concerning the fit to the Balmer jump.
The interesting case of \fnii\ is considered below.

The IR CELs merit special attention as they can provide information about the
enrichment of Ne, S and Ar in the posited H-poor component relative to that
obtained for CNO; i.e. whether model D2 is to be preferred over D1. Out of 13
IR CELs listed in Table~3, \fnii\ 122- and \farii\ 7\,$\mu$m are extremely weak
(or undetected) and their fluxes should be considered uncertain by at least a
factor of 2. According to Vermeij et al. (2002a), the most uncertain fluxes are
those of \fsiii\ 33- and \fneiii\ 36\,$\mu$m, which both suffer from the
uncertain calibration of the SWS-4 detector (by at least 30 per cent); \fneii\
12.8\,$\mu$m (25--30 per cent accuracy); and the weak \fariii\ 22\,$\mu$m. For
the purpose of determining the IR calibration self-consistently by comparison
to model results, lines which significantly depend on physical conditions
should be discarded, and several subsets of IR CELs can be identified
(Table~4). In addition to \fsiii\ 33\,$\mu$m and, particularly, \foiii\
88\,$\mu$m, whose critical densities are low, \fsiv\ 10.5\,$\mu$m is also
strongly model-dependent in that, analogously with \fariv\ (Section~3.3.1), it
is a high-ionization line sensitive to both the adopted primary spectrum and
the density distribution. Thus, an accurate calibration could \emph{a priori}
be based on `set IR5', comprising the 5 remaining lines (i.e. IR5: \fniii\ 57-,
\foiii\ 52-, \fneiii\ 15.5-, \fsiii\ 18.6-, and \fariii\ 9\,$\mu$m). Out of
these \fniii\ is peculiar since it has no optical counterpart and its predicted
flux is again model-dependent. Accordingly, the most reliable,
model-independent set is IR4, i.e. IR5 minus \fniii\ 57\,$\mu$m. A set
including all IR CELs with reliable fluxes is IR7, i.e. IR5 with the addition
of \foiii\ 88- and \fsiv\ 10.5\,$\mu$m. Finally, IR11 is the set including the
11 IR CELs that were well detected. In Table~4 the means and 1\,$\sigma$
scatters in the departure ratios in models S1--2 and D1--2 are provided for the
4 line sets identified above.

\setcounter{table}{3}
\begin{table}
\caption{Average departure ratio and 1$\sigma$ scatter for IR lines.}
\begin{tabular}{lcccc}
\noalign{\vskip3pt} \noalign{\hrule} \noalign{\vskip3pt}
Line set   & S1            & S2            & D1            & D2           \\
\noalign{\vskip3pt} \noalign{\hrule}\noalign{\vskip3pt}

\noalign{\vskip3pt}

IR4$^a$    & 0.99 $\pm$ 0.07    & 1.00 $\pm$ 0.08    & 1.19 $\pm$ 0.13   & 1.39 $\pm$ 0.08    \\
IR5$^b$    & 1.10 $\pm$ 0.22   & 1.02 $\pm$ 0.08    & 1.29 $\pm$ 0.20   & 1.38 $\pm$ 0.07    \\
IR7$^c$    & 1.17 $\pm$ 0.24   & 1.16 $\pm$ 0.30   & 1.36 $\pm$ 0.21   & 1.37 $\pm$ 0.12   \\
IR11$^d$   & 0.99 $\pm$ 0.36   & 0.96 $\pm$ 0.44   & 1.20 $\pm$ 0.28   & 1.22 $\pm$ 0.21   \\

\noalign{\vskip3pt} \noalign{\hrule}\noalign{\vskip3pt}
\end{tabular}
\begin{description}
\item[$^a$] IR4: \foiii\ 52-, \fneiii\ 15.5-, \fsiii\ 18.6-, and \fariii\ 9\,$\mu$m.
\item[$^b$] IR5: IR4 + \fniii\ 57\,$\mu$m.
\item[$^c$] IR7: IR5 + \foiii\ 88-, \fsiv\ 10.5\,$\mu$m.
\item[$^d$] IR11: IR7 + \fneii\ 12.8-, \fneiii\ 36-, \fsiii\ 33-, and \fariii\ 21.8\,$\mu$m.
\end{description}
\end{table}

The small scatter for IR4 indicates that an IR calibration from accurately
measured lines can be reliably defined (with some reservations for D1),
provided that these lines are weakly dependent on physical conditions. This
suggests that the employed collisional/radiative atomic data and \emph{ISO}
observations are of fair accuracy, and that main features of the present models
are not grossly in error. While our initial IR calibration is suited to models
S1--2, upward corrections of the IR line intensities by 19 and 39 per cent are
indicated from D1 and D2 respectively. This shift is not a weakness of the D
models since the absolute calibration of the IR relative to the optical is
rather poor: e.g. note that the \hb\ flux, as derived from the \hi\ IR lines,
is more than twice the one estimated by Dufour et al. (1982) falling into their
\emph{IUE} 10$''$$\times$20$''$ lobe (comparison made after reduction to an
equally sized aperture; the SWS and Dufour et al. observed a very similar
position). Thus an upward calibration of the IR lines relative to the optical
serves to reduce the disagreement between these two $F$(\hb) estimates. In this
respect model D2 should be preferred.

Upon adding \fniii\ (IR5), $\sigma$ increases markedly for BB models (S1, D1),
whereas it is stable for \emph{CoStar} models (S2, D2), showing that, by
adopting the latter primary spectrum, the consistency between the optical
\fnii\ CELs (used to determine N/H) and the IR \fniii\ line is preserved.
Considering IR7, which comprises all accurately measured IR CELs, $\sigma$ is
rather stable (and large) for S1 and D1, but is dramatically increased for S2
due to the large departure ratio of \fsiv. In contrast to S2, model D2 is only
mildly affected thanks to the conjunction of (i) a smaller departure ratio for
\fsiv\ due to the relatively high pressure in the inner zone of this model, and
(ii) an overall enhancement of the IR line intensities (the factor of 1.39
upward shift in calibration) partly due to the enhanced abundances of Si--Fe in
the H-poor gas (see below).

The full IR11 set leads to a general decrease of the averages, with a
correlatively larger $\sigma$, partly attributed to the lesser accuracy of the
4 additional lines. Nonetheless, model D2 again fares better with a $\sigma$ of
$\sim$20 per cent, probably not much larger than the one expected \emph{a
priori} on the basis of the claimed accuracy of the employed line fluxes. The
decrease of the averages from IR7 to IR11 is mainly due to \fariii\ 21.8- and
\fneii\ 12.8\,$\mu$m, with the former being by far the weakest line of the set
and the latter strongly model-dependent, and hence relevant for model
appraisal. The particularly bad score of S1 and S2 when all IR CELs are
included is to a large extent a consequence of the weakness of the predicted
\fneii\ intensity, while the acceptable score of D1 relates to an excellent fit
to \fneii. Here the dual abundance models constitute an improvement over the
homogeneous models, with a large fraction of \fneii\ arising from the H-poor
gas. In this respect, D2 may appear somewhat less successful than D1; the fact
however that \fneii\ is underestimated in D2 does not necessarily imply a
failure of the model, as the same occurs in an unpublished {\sc nebu} model of
NGC\,7027 using the same atomic data. Assuming on this basis a 50 per cent
improvement to the \fneii\ 12.8\,$\mu$m fit and ignoring \fariii\ 21.8\,$\mu$m,
the 1\,$\sigma$ scatter in the departure ratios for the full set of IR CELs is
only 14 per cent in D2.

Concerning elements beyond magnesium and owing to the large S/H and Ar/H in the
H-poor component of model D2, the two most representative lines \fsiii\ 18.7-
and \fariii\ 9\,$\mu$m are enhanced relative to D1, and are in harmony with the
most representative IR CELs of N, O and Ne. This is the origin of the
remarkably small scatter found for IR CELs in D2 (Fig.\,~2), indicating that
elements beyond magnesium are potentially enhanced along with oxygen in the
posited H-poor gas.

Another clue to the superiority of model D2 over D1 is provided by \nii\ and
\fnii. In model D1, the temperature of the H-poor gas is not low enough to
fully quench the \fnii\ $\lambda$6584 emission, as O/H cannot be very large,
given the constraint imposed by the weak \oi\ $\lambda$7773 ORL. Since N/H
there is bound to be sufficiently large to account for the \nii\ $\lambda$5679
ORL, \fnii\ $\lambda$6584 is enhanced, which leads in turn to a decrease of N/H
in the `normal' component, with the consequence that \fnii\ $\lambda$5755,
which was somewhat overestimated in homogeneous models, is now underestimated
(this line cannot be produced in the H-poor gas, because of its high excitation
temperature). Lowering N/H then leads to an increase of N/H in the H-poor gas
in order to account for \nii\ emission, thus potentially causing a kind of
divergence of N/H in both components, with an increasingly bad fit to the
optical \fnii\ line ratio. Although this divergence does not really occur in
D1, it can explain why the predicted \fnii\ $\lambda$5755/$\lambda$6584 ratio
in D1 is significantly low (Table~3), and why the overall N/H in the same model
is anomalously small (being almost equal to that in S1, contrary to all other
abundances; see Table 2). On the other hand, the extra cooling provided by
notably \fsiii\ and \fariii\ in the H-poor component of D2 is enough to almost
quench \fnii\ $\lambda$6584 emission, thus preventing the `\nii\ divergence'
and leading to an optimal fit for the optical \fnii\ line ratio. In this sense,
the dual abundance models are generally more successful than their homogeneous
counterparts and, among them, D2 is by far the most successful.

Finally, the scatter in the IR line departure ratios in D2 can be scrutinized
from yet another standpoint. Among the 4 lines added to IR7 to make up IR11,
the very strong \fsiii\ 33\,$\mu$m stands out. Adding this line only to IR7,
$\sigma$ increases from 12 to 15 per cent. As previously noted, \fsiii\ 33- and
\foiii\ 88\,$\mu$m, are the only lines which can be seriously affected by
collisional de-excitation in the low density conditions of 30\,Dor. The average
for the remaining 6 lines of IR7 is 1.42, close to the re-calibration factor of
1.39 obtained from IR4, and the scatter is again 8 per cent only (similar to
that found for IR4 or IR5; Table 4). Thus after re-calibration, \emph{all}
strong IR CELs in D2 are predicted in remarkable agreement with observation,
with the exception of \foiii\ 88- and \fsiii\ 33\,$\mu$m, which are both
underestimated by a factor of $\sim$1.3 (Fig.\,~2), suggesting that the fields
of view of LWS and SWS-4 sampled regions of lower average density than the
bright `optical' filament. In a trial calculation with $R_{\rm in}$ twice
larger and the gas pressure one third that of model D2, the IR \foiii\
predicted line ratio was inverted, while the \fsiii\ ratio was better fitted
without being inverted; predictions for other IR CELs remained fair. This
suggests that taking into account the background emission falling into the
\emph{ISO} lobe could help in improving the fit to \foiii\ 88- and \fsiii\
33\,$\mu$m, although the \fsiii\ line ratio would still be off somewhat. At
this point we recall that the \fsiii\ 33\,$\mu$m line was not amongst the most
accurately measured, and that the collision strengths for the \fsiii\ IR CELs
are probably not of ultimate accuracy; the values by Galavis et al. (1995) used
in the present investigation, differ very substantially from those obtained by
Tayall \& Gupta (1999).

In conclusion, the few imperfections of model D2 can be naturally understood as
a consequence of the coarseness of the `one-filament approximation' and/or
uncertainties in the observations and the atomic data. The global upward
re-calibration of the IR line fluxes required by this model is warranted
because the gap between the SWS and direct estimates for $F$(\hb) is then
substantially narrowed. The dual-abundance models generally improve the fit to
several lines other than the CNO ORLs, and D2 represents the most satisfactory
compromise since the scatter in the departure ratios of IR CELs is much
reduced. An important consequence of the above discussion is that it is quite
plausible that the posited H-poor component is enriched (relative to H) not
only in CNONe, but also in heavier elements.

\subsubsection{Ionization structure}

In Fig.\,~3 comparison is shown between the ionic fractions of \np\ and \op\ in
the two gas phases of model D2, while in Figs.~\,4 and 5 the comparison
involves the ionic fractions of the doubly ionized species of CNONe in each of
the two D2 nebular components. The IR \foiii, \fneiii\ and \fsiii\ lines are
the principal coolants in the clumped phase and are emitted instead of the
optical \foiii\ coolants because they are favoured by the lower prevailing
\elt, while the clump density (\eld\ = 696\,\cmt) is not sufficient to severely
quench most of the lines. In this respect, the model is in harmony with the
suggestion made by T03 who, based on the observed deviations of \oii\ ORL
relative intensities from their theoretical LS-coupling ratios, proposed that
the ORL emission in 30\,Dor (and other \hii\ regions) originated in low density
regions and not in exceedingly dense, supposedly ionized clumps (see also Ruiz
et al. 2003).

\begin{figure}
\centering \epsfig{file=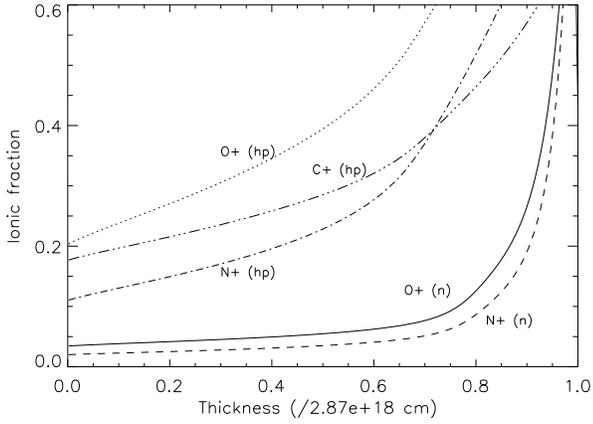, width=8.6 cm, clip=,} \caption{Ionic fractions
of singly ionized carbon, nitrogen and oxygen in the H-poor and normal gas
phases of model D2 \emph{vs.} the thickness of the model filament;
`n' and `hp' labels as in Fig.\,~1.} 
\end{figure}

\begin{figure}
\centering \epsfig{file=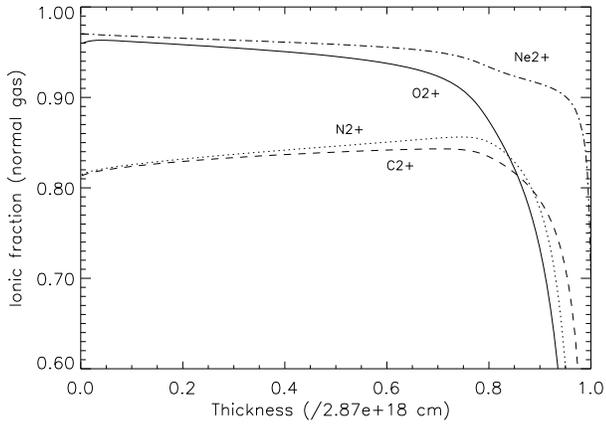, width=8.6 cm, clip=,} \caption{Ionic fractions
of doubly ionized CNONe in the normal gas phase of model D2 \emph{vs.}
thickness.}
\end{figure}

\begin{figure}
\centering \epsfig{file=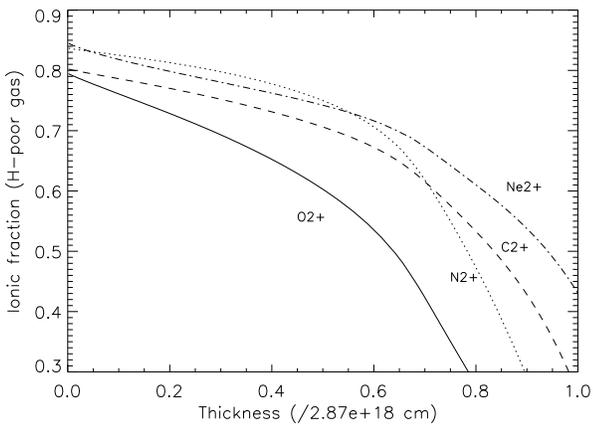, width=8.6 cm, clip=,} \caption{As in Fig.\,~4,
for the H-poor gas phase of model D2.}
\end{figure}

\begin{figure}
\centering \epsfig{file=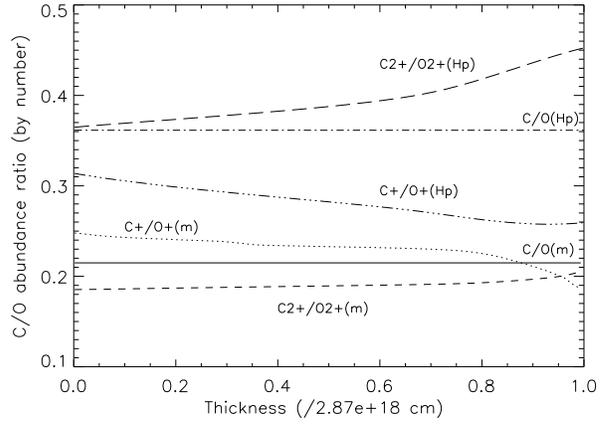, width=8.6 cm, clip=,} \caption{The C/O
elemental and ionic abundance ratios in model D2 \emph{vs.} the thickness of
model filament; curve labels marked by `m' refer to \emph{mean} quantities for
the whole nebula, while those marked by `Hp' refer to quantities within the
H-poor phase only.}
\end{figure}

\begin{figure}
\centering \epsfig{file=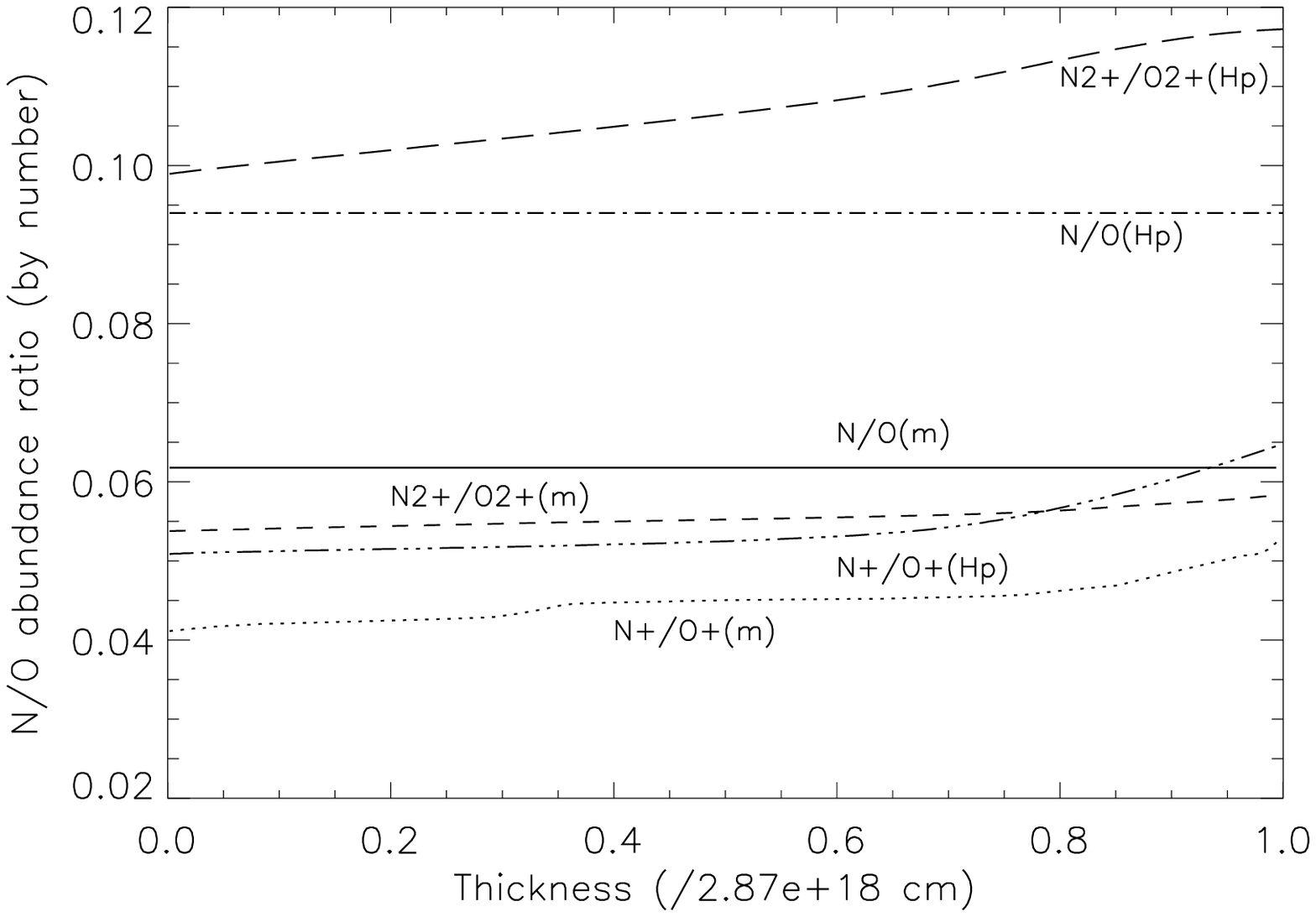, width=8.6 cm, clip=,} \caption{As in Fig.\,~6,
for N/O.}
\end{figure}

\begin{figure}
\centering \epsfig{file=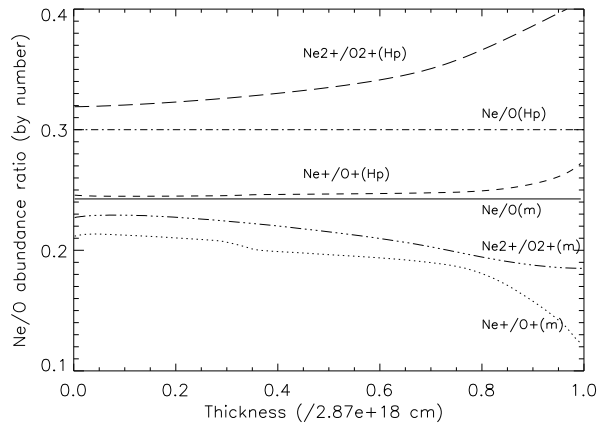, width=8.6 cm, clip=,} \caption{As in Fig.\,~6,
for Ne/O.}
\end{figure}

The variation of the ionic and elemental C/O, N/O and Ne/O abundance ratios
across the nebula was studied (Figs.~\,6--8). From a comparison with results
obtained from observations, some insight can be gained on the best way of
determining these important quantities. Here, the model C/O, N/O and Ne/O
ratios (by number) are well approximated by the abundance ratio of their
respective doubly ionized species, e.g. N/O $\simeq$ \npp/\opp\ ($\neq$
\np/\op). This is true for the \emph{mean} abundance ratios for the whole
nebula, as well as for the ratios within the relatively H-poor phase only, and
is due to the high excitation state of the gas. The \np/\op\ ratio, often used
as an approximation to the N/O by observers who sample the optical \fnii\ and
\foii\ CELs only, is about two thirds the mean elemental ratio, while in the
clumps it is lower by a factor of two than the respective N/O of the H-poor
phase. These results reproduce the behaviour of N/O obtained by T03 (cf. their
Table 12) based on pure optical and pure IR forbidden-line ratios. As was also
noted by Rubin et al. (1988), Simpson et al. (1995) and T03, in several \hii\
regions IR FS line ratios yield a consistently higher N/O than optical line
ratios. According to the models shown here the IR FS lines of \fniii\ and
\foiii\ give a better estimate of the N/O ratio than the optical \fnii\ and
\foii\ CELs, both for the nebula as a whole, as well as for the posited,
embedded H-poor clumps also (if one was to image the clumps in IR nitrogen and
oxygen fine-structure lines to derive their N/O ratios). It should be mentioned
that the same behaviour of the C, N, and Ne ratios relative to oxygen is
obtained for the chemically homogeneous model S1.

\setcounter{table}{4}
\begin{table*}
\begin{center}
\caption{Temperatures (in K), \tsq\ factors and mean neutral and ionic
fractions, for the chemically homogeneous model S1, and the dual-abundance
model D2 (for the whole nebula and each of the two components) for H, He,
CNONe, and S. The numbers in parentheses are powers of 10.}
\begin{tabular}{lccccccccc}
\noalign{\vskip3pt} \noalign{\hrule} \noalign{\vskip3pt}
Element   &   &{\sc i}    & {\sc ii}     &{\sc iii}       &  {\sc iv}                   &{\sc i}    & {\sc ii}     &{\sc iii}       &  {\sc iv}       \\
\noalign{\vskip3pt} \noalign{\hrule} \noalign{\vskip3pt}
&\multicolumn{6}{c}{S1}                                                                     &\multicolumn{3}{l}{~~~~~~~~~~~~~~~~~D2}               \\
H & (\tion)     & 9958        &9962            &                &                    &10920      &9654       &          &                        \\
  & (\tsq)      & 1.29($-$2)   &4.45($-$3)       &                &                &2.09($-$2)  &2.23($-$2)  &          &                              \\
  &(X$^{+i}$/X) & 9.66($-$3)   &9.90($-$1)       &                &                &8.03($-$3)  &9.92($-$1)  &          &                       \\
He     &    &10150        & 9960           & 9807             &                      &9741       &9615       &9174       &                        \\
       &    &1.28($-$2)    & 4.45($-$3)      & 1.65($-$5)        &                &6.72($-$2)  &2.52($-$2)  &1.70($-$2)  &                            \\
       &    &9.35($-$3)    & 9.91($-$1)      & 1.57($-$4)        &                &7.70($-$3)  &9.92($-$1)  &2.92($-$4)  &                          \\
C      &    &10860        & 10610          & 9904             & 9592                 &5421       &6585       &7821       &9089                    \\
       &    &3.76($-$3)    & 5.83($-$3)      & 3.92($-$3)        & 1.03($-$3)    &2.11($-$1)  &2.28($-$1)  &1.44($-$1)  &2.84($-$2)                \\
       &    &3.58($-$5)    & 9.92($-$2)      & 8.66($-$1)        & 3.45($-$2)    &1.06($-$4)  &1.99($-$1)  &7.47($-$1)  &5.31($-$2)                      \\
N      &    &9433         & 10790          & 9876             & 9600                 &7162       &7129       &7901       &8789                    \\
       &    &2.09($-$2)    & 4.68($-$3)      & 3.58($-$3)        & 1.12($-$3)    &1.75($-$1)  &2.07($-$1)  &1.37($-$1)  &5.17($-$2)                \\
       &    &2.49($-$3)    & 1.07($-$1)      & 8.53($-$1)        & 3.73($-$2)    &1.60($-$3)  &1.91($-$1)  &7.40($-$1)  &6.69($-$2)                  \\
O      &    &9868         & 10730          & 9818             & 9808                 &9570       &7533       &8679       &8425                    \\
       &    &1.33($-$2)    & 4.58($-$3)      & 3.06($-$3)        & 2.03($-$5)    &9.45($-$2)  &1.96($-$1)  &7.78($-$2)  &7.37($-$2)                 \\
       &    &8.66($-$3)    & 1.58($-$1)      & 8.33($-$1)        & 6.48($-$5)    &6.07($-$3)  &2.30($-$1)  &7.64($-$1)  &7.10($-$5)                \\
Ne     &    &9172         & 10380          & 9901             & 9797                 &5363       &5843       &8046       &8031                   \\
       &    &2.22($-$2)    & 5.58($-$3)      & 4.02($-$3)        & 2.45($-$5)    &1.40($-$1)  &2.27($-$1)  &1.30($-$1)  &1.04($-$1)               \\
       &    &1.33($-$3)    & 1.28($-$1)      & 8.71($-$1)        & 6.45($-$6)    &8.13($-$4)  &1.44($-$1)  &8.56($-$1)  &1.56($-$6)                \\
S      &    &10270        & 10550          & 10130            & 9600                 &6885       &8134      & 8348       &9152                          \\
       &    &1.03($-$2)    & 7.64($-$3)      & 4.90($-$3)        & 1.06($-$3)    &2.38($-$1)  &1.70($-$1) & 1.20($-$1)  &2.40($-$2)                   \\
       &    &2.46($-$6)  & 3.31($-$2)      & 6.31($-$1)        & 3.21($-$1)      &3.51($-$6)  &5.49($-$2) & 6.72($-$1)  &2.52($-$1)                 \\

\noalign{\vskip3pt}
&\multicolumn{6}{c}{D2 (H-poor gas)}                                                                       &\multicolumn{3}{l}{~~~~~~D2 (Normal gas)}        \\
H & (\tion)   &5325         &4120            &                &                     &10500       &9897        &          &                 \\
  & (\tsq)    &3.70($-$2)    &6.29($-$2)       &                &                     &6.71($-$3)   &7.20($-$3)   &          &                 \\
  &(X$^{+i}$/X) & 3.04($-$2) &9.70($-$1)       &                &                     &1.80($-$2)   &9.82($-$1)   &          &                  \\
He     &      &5158         &4107            &3128              &                   &10370       &9903        &9407       &                 \\
       &      &4.48($-$2)    &6.24($-$2)       &1.51($-$4)         &                   &7.83($-$3)   &7.23($-$3)   &2.70($-$5)  &                 \\
       &      &4.79($-$2)    &9.52($-$1)       &2.05($-$4)         &                   &1.13($-$2)   &9.88($-$1)   &2.90($-$4)  &                 \\
C      &      &4699         &4672            &3847              &3421               &10740       &10550       &9850       &9465               \\
       &      &5.42($-$2)    &5.74($-$2)       &5.17($-$2)         &2.31($-$2)          &6.63($-$3)   &7.85($-$3)   &6.68($-$3)  &2.43($-$3)         \\
       &      &2.56($-$4)    &3.80($-$1)       &6.12($-$1)         &7.89($-$3)          &2.85($-$5)   &1.26($-$1)   &7.94($-$1)  &7.95($-$2)          \\
N      &      &5304         &4875            &3718              &3376               &10350       &10840       &9800       &9484              \\
       &      &3.62($-$2)    &4.78($-$2)       &4.10($-$2)         &1.84($-$2)          &8.26($-$3)   &6.89($-$3)   &5.96($-$3)  &2.63($-$3)         \\
       &      &1.53($-$2)    &3.64($-$1)       &6.01($-$1)         &1.97($-$2)          &2.21($-$3)   &1.30($-$1)   &7.78($-$1)  &9.04($-$2)         \\
O      &      &5510         &4624            &3601              &3131               &10460       &10790       &9736       &9407               \\
       &      &2.67($-$2)    &5.38($-$2)       &3.45($-$2)         &1.48($-$4)          &6.45($-$3)   &7.38($-$3)   &5.35($-$3)  &5.11($-$5)          \\
       &      &2.81($-$2)    &4.91($-$1)       &4.81($-$1)         &4.07($-$5)          &1.59($-$2)   &1.52($-$1)   &8.32($-$1)  &7.85($-$5)         \\
Ne     &      &5175         &4587            &3946              &3129               &9864        &10250       &9885       &9393           \\
       &      &4.61($-$2)    &5.93($-$2)       &5.81($-$2)         &2.88($-$4)          &7.00($-$3)   &8.11($-$3)   &7.11($-$3)  &3.95($-$5)       \\
       &      &6.43($-$3)    &3.17($-$1)       &6.77($-$1)         &8.48($-$7)          &6.15($-$4)   &6.26($-$2)   &9.37($-$1)  &1.91($-$6)      \\
S      &      &4839         &4933            &4070              &3378               &10590       &10740       &10080      &9465         \\
       &      &5.55($-$2)    &5.24($-$2)       &5.95($-$2)         &1.96($-$2)          &6.61($-$3)   &6.64($-$3)   &7.71($-$3)  &2.32($-$3)   \\
       &      &1.23($-$5)    &1.43($-$1)       &8.05($-$1)         &5.22($-$2)          &3.00($-$6)   &5.22($-$2)   &6.16($-$1)  &3.05($-$1)    \\

\noalign{\vskip3pt} \noalign{\hrule} \noalign{\vskip3pt}
\end{tabular}
\end{center}
\end{table*}

\subsection {On the \tsq\ factors from models}

Modelled mean temperatures (\tion), mean squared temperature fluctuation (\tsq)
factors (Peimbert 1967) and ionic fractions (X$^{+i}$/X) are shown in Table~5
for the whole shell of model D2 (upper right panel), the H-poor component
(lower left) and the normal component (lower right) for the elements H, He,
CNONe and S; \tion\ and \tsq\ values were derived using, e.g. equations 3 \& 4
of P03. Under the assumption for the existence of temperature fluctuations in a
chemically homogeneous nebular medium, the elemental abundances derived from
CELs overall underestimate the true abundances due to the adoption of a
non-representative (overestimated) electron temperature for all nebular zones.
Under this premise the heavy-element ORLs, which in most cases yield higher
abundances than CELs, are often thought to give a more accurate measure of the
nebular abundances since their strengths relative to \hb\ are much less
dependent on \elt\ (and \eld) than any given CEL to \hb\ line ratio. Because of
this, various workers have adopted a methodology whereby abundances of elements
from carbon to iron, and derived from CELs, are often corrected upwards. The
corrections are estimated by adopting a representative \tsq\ factor which is
usually the average of those computed from the difference between the Balmer
discontinuity and \foiii\ forbidden-line temperatures (\tsqb), and those
computed from the ORL vs. CEL abundance discrepancy factor (\tsqd; e.g. from
\cpp and \opp\ ions). P03 adopted \tsq $=$ 0.033 for 30\,Dor corresponding to
\emph{rms} temperature fluctuations across the nebula of 18\,per cent. Similar
values were deduced from the \opp\ ADFs by Peimbert (2003) and T03. Slightly
higher values were found from the \cpp\ ADF by the same authors. Our modelling
results however allow for an alternative interpretation of the observations.

The single component, chemically homogeneous model S1 yields \tsq(\hp) $=$
0.0045, i.e. the \emph{rms} \elt\ fluctuations over the whole \hii\ zone are
7\,per cent (Table~5). Previous work on photoionization nebular models under
the assumption of chemical homogeneity and variable gas density did not reveal
\tsq\ factors much in excess of $\sim$0.01 for a wide range of ionizing colour
temperatures and densities (Kingdon and Ferland 1995; hereafter KF95). In model
S1 the \tsq\ factors for doubly ionized CNONe are all less than \tsq(\hp) as
they should be; this is a well known result (Harrington et al. 1982; KF95)
arising from the non-coincidence of the \hp\ and \opp\ zones. KF95 warn against
a direct comparison between \tsqb\ and model-derived \tsq\ factors, such as
those listed in Table~5, because the comparison is valid only when the mean
ionic temperatures and \tsq\ factors for \hp\ and \opp\ are equal. This is not
exactly true even for our chemically homogeneous model S1 where
\tion(\opp)$-$\tion(\hp) $=$ $-$144\,K and \tsq(\opp) $=$ 0.69\tsq(\hp). For
the dual abundance model D1, these values are respectively $-$975\,K and
3.5\tsq(\hp). P03 found \tsqb $=$ 0.022, interpreting this result as evidence
for temperature fluctuations in a chemically homogeneous nebula. This is not
supported by model S1. It should further be noted that both S1 and D1 models
fit the Balmer discontinuity within a 3\,per cent margin (as well as the
optical \foiii\ lines). Only models D1--2 however reproduce the intensities of
the CNO ORLs. For model D2 globally, \tsq(\hp) $=$ 0.022, \tsq(\opp) $=$ 0.078
and \tsq(\cpp) $=$ 0.14; thus the introduction of H-deficient inclusions
results in the derivation of higher \tsq\ values that approach or exceed the
(empirical) \tsqb. Even so, this verisimilitude cannot be interpreted in the
framework of the original \elt-fluctuation hypothesis since the empirical
\tsqb\ has no direct physical correspondence in the context of these models.
This is also evident from Fig.\,~1 where the electron temperature profile
across the physical extent of the two gas phases of model D2 is plotted. A
substantial temperature gradient can only be discerned throughout the H-poor
phase whose \elt\ rises, almost monotonically, from $\sim$3000\,K at the inner
edge to 6200\,K at the outer model radius. Similarly, the empirical \tsqd\
values cannot be meaningfully compared with the model's \tsq\ values for
various ions.

\subsection {Model \emph{versus} empirical \tsq\ elemental abundances}

\setcounter{table}{5}
\begin{table*}
\centering
\begin{minipage}{120mm}
\caption{Gas-phase abundances (by number) from single and dual abundance models
compared to empirical abundances, in units such that $\log$$N$(H) $=$ 12.}
\begin{tabular}{lccccccccc}
\noalign{\vskip3pt} \noalign{\hrule} \noalign{\vskip3pt}
        &S1    &S2     &D1      &D2    &This work   &P03$^a$  &P03$^a$   &T03$^b$   &V02$^c$  \\
        &      &       &Mean    &Mean  &adopted  &\tsq\ $=$ 0.0 &\tsq\ $=$ 0.033 &\tsq\ $=$ 0.0  &\tsq\ $=$ 0.0\\
\noalign{\vskip3pt} \noalign{\hrule} \noalign{\vskip3pt}
He      &10.93 &10.93  &10.92   &10.92    &10.92      &10.94  &10.93     &10.96     &10.97 \\
C       &7.64  &7.66   &7.78    &7.78     &7.78       &7.80    &8.05     & --       & --   \\
N       &7.24  &7.15   &7.25    &7.24     &7.24       &7.05    &7.21     & 6.89     & 6.84 \\
O       &8.36  &8.37   &8.45    &8.45     &8.45       &8.33    &8.54     & 8.34     & 8.25 \\
Ne      &7.64  &7.60   &7.73    &7.73     &7.73       &7.65    &7.83     & 7.66     & 7.84 \\
Mg      &6.72  &6.60   &6.74    &6.65     &6.69       & --     & --      & --       & --  \\
Si      &6.40  &6.45   &6.46    &6.53     &6.50       & --     & --      & --       & --  \\
S       &6.79  &6.83   &6.83    &6.85     &6.84       &6.84    &6.99     & 6.77     & 6.82 \\
Cl      &4.77  &4.84   &4.82    &4.90     &4.87       &4.75    &4.82     & 4.85     & --   \\
Ar      &6.08  &6.10   &6.12    &6.16     &6.14       &6.09    &6.26     & 6.15     & 6.21 \\
Fe      &6.07  &6.04   &6.14    &6.11     &6.12       &6.25    &6.39     & --       & --   \\
\noalign{\vskip3pt} \noalign{\hrule} \noalign{\vskip3pt}
\end{tabular}
\begin{description}
\item[$^a$] Peimbert (2003).
\item[$^b$] Tsamis et al. (2003a).
\item[$^c$] Vermeij et al. (2002b).
\end{description}
\end{minipage}
\end{table*}

The abundances derived from our 4 models are presented in Table~6 (cols. 2--5),
along with our adopted preferred estimates (col. 6; mean of models D1 and D2)
and those from previous determinations (cols. 7--10).  They do not include
estimates for depletion onto dust grains, so they pertain to the gaseous phase
only. Both the \tsq\ $=$ 0.0 and \tsq\ $=$ 0.033 results of P03 are listed (the
latter were favoured by P03). Our preferred abundances are lower overall than
those for \tsq\ $=$ 0.033, but nitrogen is 10\,per cent larger; this is partly
due to the coarse correction for unseen ions applied by P03 (see Sect.~3.4.3).
The differences are largest, about a factor of 2, for C and Fe. These are the
elements which are mostly affected by the upward correction under the
assumption of \tsq\ = 0.033.

At this point it should be noted that abundances determined from detailed
photoionization models and empirical methods cannot be immediately ranked.
Models have the obvious advantage of self-consistency and of optimized
(although possibly wrong) ionization correction factors, whereas empirical
methods exactly `fit' diagnostic lines by definition. The presented models
exactly fit all the standard diagnostic lines. While this accuracy is illusory
for obvious reasons, it does ensure that, for given atomic data, the model
abundances arguably supersede the empirical ones.

The \tsq\ $\neq$ 0 empirical approach not only supersedes the classical one
(\tsq\ $=$ 0) in that it integrates new spectral features, but it also takes
advantage of fitting approximately the heavy element ORLs to claim superiority
over models which do not fit them, despite the fact that there is no strong
physical basis for the existence of large amplitude temperature fluctuations in
chemically homogeneous nebulae.\footnote{It should be noted, however, that
Stasi\'{n}ska \& Szczerba (2001) proposed nebular models which resulted in
enhanced \tsq\ values by considering small scale density variations in a dusty
medium.} Here we exhibit photoionization models based on known standard
physics, which account for the ORLs as well as for other optical spectroscopic
diagnostics thanks to the fundamental assumption of chemical inhomogeneity,
which we consider to be plausible on independent grounds (Section~4). In
addition, despite the demanding condition of physical self-consistency, the
models account well for a number of independent observables, while the
remaining moderate discrepancies are understood in terms of identified
uncertainties and the coarseness of the single filament description. Therefore
even though the models are probably not unique, their (averaged) abundances
provide a basis on which empirical \tsq\ abundances can be evaluated in this
particular situation.

A comparison of the results is meaningful if differences in the employed atomic
data and possible inaccuracies in the ionization correction factors adopted in
the empirical approach are neutralized. This can be achieved by considering the
results \emph{differentially}. For this purpose, the \tsq\ $=$ 0 abundances
obtained by P03 can be associated to models S1--2 (fit to optical/UV CELs), and
the \tsq\ $=$ 0.033 abundances to models D1--2 (fit to both optical/UV CELs and
ORLs). Excluding helium and elements not considered by P03, the average
abundance ratios of D1 over S1 and D2 over S2 are both equal to 1.20 $\pm$ 0.10
(the aberrant N abundance of D1 was discarded; see discussion on \fnii\ in
Sect.~3.4.2), while the same ratio for the `\tsq\ $=$ 0.033' over `\tsq\ $=$ 0'
abundances is 1.48 $\pm$ 0.18. Although we have no clear explanation of why
this is so, we note that the \emph{square root} of the `\tsq\ abundance ratio',
1.21 $\pm$ 0.07, is identical to the `D/S model ratio' (On close inspection of
Table~6, however, the individual abundance enhancements differ significantly in
the two approaches). Thus, compared to this particular self-consistent physical
model which quantitatively accounts for observation, the `\tsq\ $=$ 0.033'
abundances are significant overestimates, while the mean D1--2 model abundances
are equal on average to the geometric mean of the \tsq\ $=$ 0 and non-0
empirical abundances. In other words, \emph{there is objectively no reason in
this example to prefer one or the other set of empirical abundances}. On the
other hand, concerning physical models, it is noted that whereas a model which
fits existing observational data might be wrong, one which does not almost
certainly is. On these grounds we deem that adopting average abundances from
models D1--2 as representative is a justified improvement compared to values
derived from either chemically homogeneous models or
empirical methods. \\

\subsection {On the composition of the H-poor gas}

\subsubsection {Helium and the primordial He abundance}

The question of whether He is enriched in the clumped gas along with CNONe is
of interest, since helium is expected to be present in e.g. nucleosynthetic
products expelled from evolved, massive WR-type stars (e.g. Rosa \& Mathis
1987; Kobulnicky et al. 1997). The larger the helium abundance is in the
H-deficient gas, the smaller it is in the `normal' nebular component. A recent
value for the primordial He/H is about 0.0786 by number (Luridiana et al.
2003). From model trials, the condition that He/H $>$ 0.08 in the normal
component translates into He/H $<$ 0.12 for the posited H-poor clumps in
30\,Dor. Thus helium is not much enhanced relative to hydrogen in the clumps,
and the excess heavier elements there should originate in extreme
nucleosynthesis of the kind expected from supernova explosions (e.g. Woosley et
al. 2002), rather than from Wolf-Rayet stellar winds (see Section 4).

In both D1--2, He/H $=$ 0.10 , was arbitrarily adopted for the H-poor component
(between the extremes of 0.08 and 0.12; Table~2). Decreasing He/H in the clumps
below 0.10, an increase of the \emph{overall} He/H abundance follows until the
latter becomes approximately equal to the one found from chemically homogeneous
models (or standard empirical analyses). If, reasonably, it is assumed that
He/H in the H-poor gas is at least as large as in the normal gas, it naturally
follows that the He/H ratio derived from standard analyses of \hii\ regions,
under the assumption of nebular chemical homogeneity, is only an \emph{upper
limit to the actual He abundance}. A similar point was made by P\'equignot et
al. (2002) in the case of PNe. This systematic effect is a possible new source
of uncertainty in the important quest for the primordial He abundance from
GHIIR spectroscopy of blue compact galaxies (e.g. Luridiana et al. 2003; Izotov
\& Thuan 2004).

\subsubsection {Elements beyond helium}

While the major part of oxygen and other heavy elements, such as Ne, Mg, Si, S,
Cl, and Ar, is ejected into the ISM during the final stages of massive star
evolution ($M$ $>$ 8\,M$_{\odot}$) and the resulting type II supernova
explosions, the exact contribution of massive stars vs. low- and
intermediate-mass stars in the production of C and N is under debate: e.g.
Carigi et al. (2005) advocate Galactic chemical evolution models according to
which only about half of the carbon found at present in the solar vicinity has
been produced by massive stars. SN ejecta are strongly deficient in hydrogen
and constitute promising sources of `oxygen-rich clumps' such as those favoured
in the present work (see also Section~4). Are then the CNO(Ne) abundances of
the H-deficient clumps discussed herein in keeping with those expected from
massive star nucleosynthesis? First, it must be noted that the photoionization
model abundances for such H-poor clumps are likely to be affected by large
uncertainties. Concerning nitrogen in particular, only a few weak \nii\ ORLs
are available, which can result in an overestimated N/O in cases of unforseen
blends with even weaker lines; in addition, the \nii\ lines belonging to
multiplet V3 are known to be potentially partly excited by continuum
fluorescence, not only by recombination (see e.g. T03); this also points
towards an upper limit N/O ratio for the H-poor gas. The weakness of the \nii\
ORLs certainly implies that the N/O of the posited clumps is much less than
unity, but it does not provide a precise value or even a secure lower limit.
More generally, uncertainties can also arise from the fact that the abundances
of the clumps were obtained from recombination lines \emph{after} subtracting
the contribution to their emission from the `normal' composition nebular gas
(40--50, 50--70 and 65--70 per cent for \cii, \nii\ and \oii\ ORLs
respectively). Thus, while there is absolutely no doubt that the observed
intensities of at least the \cii\ and \oii\ ORLs are larger than those obtained
from chemically homogeneous models,\footnote{As mentioned in Section~1 this is
amply documented in empirical analyses of many \hii\ regions (Peimbert et al.
1993; Esteban et al. 2005), including 30 Dor (T03; P03).} the exact excess,
here some 55, 40 and 30 per cent for \cii, \nii\ and \oii\ respectively, is not
very accurately determined due to observational uncertainties. Another source
of uncertainty in the resulting abundances is the coarseness of the sampling of
the material effectively observed, given the necessarily schematic geometry of
the dual-abundance photoionization model and the fact that spatially-resolved
observations of the posited H-poor clumps are not available yet; the latter
goal might some day be achieved via means of integral-field unit spectroscopy
with very large telescopes.

If we consider that our model H-poor clumps would in reality correspond to a
homogeneous mixture of extremely H-deficient material to which `normal' ISM gas
has been partially introduced, the (corrected) relative abundances in the model
clumps of (C/O, N/O) $\sim$~(0.36, 0.09) (model D2; Table 2) can be compared to
the (C/O, N/O) $\sim$~(0.20, 0.03) ratios expected from typical massive star
nucleosynthesis for the presupernova wind \emph{plus} type II SN ejecta (cf.
fig.\,~27 of Woosley et al. 2002). The fact that the photoionization models
tend to indicate C/O and N/O ratios 2--3 times larger than `expected' may be a
potential difficulty for either the model assumptions or the proposed origin of
the posited clumps, and requires further analysis: the details of the mixing
process of H-poor with H-rich gas should especially be investigated. The same
conclusion applies to Ne/O, although in this case only upper limits to the
\neii\ ORLs are available and Ne/O relies on IR CELs, notably the \fneii\
12.8-$\mu$m transition (25--30 per cent flux accuracy). Nonetheless, keeping in
mind the aforementioned uncertainties, the departures between model results and
massive star nucleosynthetic theory are not large enough to jeopardize the
basic features of the model. The N/O ratio is found to be definitely very small
in the H-poor gas and therefore satisfactory at the semi-quantitative level
currently attainable, even for cases where part of the nitrogen that was
ejected during the presupernova evolution might not reside in oxygen-rich
clumps. On the other hand, the determination of C/O in the ISM of the LMC
involves the difficult calibration of the \fciii\ $\lambda$1909 intensity
relative to the optical, which in turn can impact the accuracy of the estimated
contribution of the `normal' gas to the emission of \cii\ $\lambda$4267.
Finally, the photoionization model analysis indicates that the IR lines are
overall best accounted for by model D2 in which elements heavier than neon,
such as the essentially better constrained sulfur and argon, are enhanced in
the H-poor gas by similar amounts to oxygen (Sect.~3.4.2); this is expected for
material akin to type II SN ejecta.

\section{Discussion and conclusions}

Self-consistent photoionization models of the giant 30\,Dor nebula in the LMC
have been constructed, precisely fitting all the optical spectroscopic plasma
diagnostics available for a bright filament observed from UV to radio
wavelengths. A dual-abundance model (D2) is favoured which incorporates small
scale chemical inhomogeneities in the form of relatively H-deficient inclusions
embedded in gas of typical LMC composition, and subjected to a \emph{CoStar}
model atmosphere ionizing spectrum. The model filament, whose gas pressure
increases outward by a factor of 2, is essentially radiation bounded and its
geometrical thickness and distance from the central source are in good
agreement with what is suggested from \hb\ images of the nebula. The relatively
cool and dense inclusions are uniformly distributed throughout the filament and
are in local pressure equilibrium with their surroundings, representing 2 per
cent of the nebular mass, while emitting about 8 per cent of the total \hb\
flux. This model, designed to closely fit the CNO ORLs (in addition to the
UV/optical CELs), leads to significantly improved predicted intensities for a
number of other optical and IR lines, which were not used in the convergence
process.
The helium abundance relative to hydrogen is about normal in the inclusions,
while all other elements for which enough information is available (C, N, O,
Ne, S, and Ar) have abundances that are enhanced by typically one order of
magnitude compared to their respective values in the ambient \hii\ region. To
our knowledge, this is the first time that an \hii\ region model is exhibited
which, incorporating only standard, dust-free physics for photoionized nebulae,
provides a satisfactory solution to the extensively researched issue of
discordant ORL vs. CEL heavy-element abundances in this class of objects.


The abundances of heavy elements in the chemically inhomogeneous 30\,Dor models
were found to be lower than those advocated by authors who consistently apply
upward corrections to the forbidden-line abundances of \hii\ regions by
postulating temperature fluctuations in a chemically homogeneous medium. For
this particular nebula, the presence of H-deficient inclusions warrants an
overall upward revision of the heavy element abundances by at most 0.1\,dex,
compared to empirical methods or models that do not take into account the heavy
element ORLs, and indicates that the latter techniques provide only an upper
limit to the helium abundance in \hii\ regions. Inasmuch as, in the current
paradigm, the recombination lines do not provide an accurate estimate of the
\emph{overall} heavy-element to H abundance ratios in a nebula, our findings
cast new light on the use of CNONe ORLs as tracers of abundance gradients
across galactic disks such as those exhibited by Esteban et al. (2005). It
could prove worthwhile to investigate these gradients further in the context of
dual-abundance \hii\ region models.

Notwithstanding the positive aspects of the presented models, we must ask
ourselves: is there any basis to justify the existence of H-deficient
inclusions in 30\,Dor or other \hii\ regions and what could their origins be?
It is often argued that since the ISM appears to be chemically homogeneous
(e.g. Moos et al. 2002; Andr\'e et al. 2003), extensive mixing of the various
chemical species should somehow occur. However, observations establish
homogeneity by comparing different sightlines over typical ISM depths of
$\sim$1\,kpc, many orders of magnitude larger than the inclusions considered
here. In fact, the 0.5\,kpc local bubble appears to be oxygen deficient
relative to the average few kpc neighbourhood (Andr\'e et al. 2003). Our
interpretation of 30\,Dor as a nebula which appears to be homogeneous on a
large scale, but is chemically \emph{inhomogeneous} at the smallest sub-parsec
sized scale, resonates with the findings of de Avillez \& Mac Low (2002) who
studied the mixing of chemical species in the ISM in the context of
supernova-driven models and concluded that: `even if the major fraction of the
ISM gas is completely mixed, lower level inhomogeneities will still be seen as
abundances are measured more carefully'. We take it that the recent wealth of
abundance determinations in nebulae involving optical recombination lines, and
extensively quoted in this work, have indeed increased the level of accuracy
with which such studies can be done. Those authors established that the
\emph{smallest} ISM mixing timescale, corresponding to the largest assumed SN
explosion rate, is about 120\,Myr, with weak abundance inhomogeneities taking
longer times to be erased than strong ones due to an exponentially declining
mixing rate for weaker inhomogeneities. As they noted however, this
representative timescale is longer than the time interval between type II SNe
and other polluting factors, such as e.g. H-deficient winds from evolved
massive stars or star formation debris, which in the meantime introduce fresh
inhomogeneities into the ISM; it is also longer than the typical lifetime of
\hii\ regions like 30\,Dor and their embedded stellar OB associations.
Therefore, according to the study of de Avillez \& Mac Low, the ISM may never
reach homogeneity under such conditions.

In an attractive scenario, Tenorio-Tagle (1996; hereafter TT96) proposes that,
after a long journey in a hot coronal phase ($T$~$\geq$~10$^6$\,K), metal-rich
material ejected by supernovae (SNe) erupting within OB associations will
eventually fall back on to the galactic disk as a rain of cool, dense
`droplets' of final typical size of $\sim$10$^{17}$\,cm and densities of
$\sim$10$^3$\,\cmt. These will be incorporated quite evenly into the atomic H
and molecular H$_2$ content of the disk, without significant mixing, until a
next generation of massive stars photoionizes the gas and triggers complete
mixing. Thus according to TT96, chemical homogeneity in a galactic disk is not
instantaneously achieved soon after the end of a starburst event, but results
as a consequence of the birth of the next massive star generation. It is
tempting to identify the remains of the droplets singled out by TT96 with the
H-deficient clumps introduced in the present photoionization models, especially
as their size, a few percent of the size of a complete photoionized filament,
and expected densities, ideally fits our needs. According to this depiction,
one can envisage a galaxy's ISM abounding with small-scale chemical
inhomogeneities which are only observed emitting metallic recombination lines,
such as \cii\ $\lambda$4267, upon their photoionization in the vicinity of a
young stellar cluster. In this context, we note that whereas the R136 cluster
within 30\,Dor is of young age (1--5\,Myr; Massey \& Hunter 1998) and only few
SNRs have been unambiguously identified in its wider vicinity (Lazendic et al
2003; Chu et al. 2004), the older association Hodge 301 located 3$'$ to the
northwest, may have hosted $\sim$40 SNe since its formation 20--25\,Myr ago
according to estimates by Grebel \& Chu (2000).

As emphasized by Scalo \& Elmegreen (2004), a general problem with timescale
arguments pertaining to ISM mixing, is that they apply to turbulent transport
(e.g. Roy \& Kunth 1995), and \emph{not} to chemical homogenization. For
genuine mixing to be effectively achieved, viscosity and molecular diffusion
are ultimately required, but these mechanisms act over very small scale lengths
and/or very long times. Molecular diffusion becomes effective in the presence
of strong gradients of concentration, e.g. at the interface of two media with
very different composition. Nonetheless, the \emph{different thermal behaviour}
of such media could also delay the mixing. This is a very complex problem for
which a generic solution has yet to be found. Along the same lines, the
detailed simulations of de Avillez \& Mac Low (2002) point to a very long
timescale, over 10$^8$\,yr, for the final decay of small ISM inhomogeneities.
Future work may involve a coupled hydrodynamics/photoionization computational
approach to this problem (e.g. Lim \& Mellema 2003).

In the last step of the scenario developed by TT96, the disruption of the
H-poor droplets by photoionization proceeds hydrodynamically aided by the
presence of shocks, and the gas, now at 10$^4$\,K, acquires densities similar
to those of the ambient \hii\ region; diffusion of the disrupted droplets is
then suddenly accelerated. TT96 estimated a final diffusion time of the order
of 10$^4$\,yr in the lowest density \hii\ region gas (at the small spatial
scale of the metal-rich droplets), and a time significantly shorter than
10$^6$\,yr in denser filaments. At first view, this timescale for complete
chemical homogenization in the newly born \hii\ region seems uncomfortably
short to justify identification of the droplets with the H-poor clumps posited
in the present photoionization models. Importantly however, we note that,
unlike the supposition made by TT96, even in the event of its re-ionization,
the H-poor gas can \emph{maintain} a very low temperature due to fine-structure
line cooling and \emph{remain} relatively dense and isolated from its
surroundings for some time. Thus, considering that the thermal properties of
the H-poor inclusions will tend to significantly delay diffusion, it is quite
possible that chemical inhomogeneities will survive for a significant fraction
of the lifetime of a GHIIR. Under these conditions, identification of our model
H-deficient clumps with the relics of T96's metal-rich droplets appears to be a
reasonable guess. Additional work will be needed to develop observational tests
of the model and to further constrain the abundances in the oxygen-rich
inclusions. Whether a significant amount of nitrogen is present or not in the
H-poor phase is critical with regard to the latter's origin in material lost
e.g. during the presupernova evolution of massive stars or in a type II
supernova event (Section 3.7.2). Confirming with improved accuracy that
elements beyond neon are overabundant in the clumps in the same way as oxygen
(Section 3.4.2), is equally important.

In conclusion, the photoionization model description proposed here, which
primarily aimed at quantitatively solving a long-standing spectroscopic anomaly
using known, conventional \hii\ region physics, also turns out to be in
qualitative agreement with independent theoretical considerations regarding the
chemical evolution of the ISM. Detailed studies of other young, relatively
metal-poor GHIIRs, such as 30\,Dor, may be a promising way to unravel
small-scale composition inhomogeneities and provide additional observational
support to the interesting thesis advocated by TT96. Efforts should be made for
the detection of ORLs such as \cii\ $\lambda$4267 and \oii\ $\lambda$4650 (and
others of nitrogen and neon ions) in \hii\ regions of all kinds, in order to
build abundance determinations on firm grounds. Whether chemical inhomogeneity
will prove to be the rule or the exception is an important question in its own
right.

We note that the survival of hydrogen-deficient inclusions in the ISM could be
of paramount importance for the process of star formation since these
inclusions, cooler and denser than their surroundings, could act as seeds to
trigger protostellar collapse.

\vspace{7mm} \noindent {\bf Acknowledgments}

YGT acknowledges support from a Peter Gruber Foundation Fellowship awarded by
the IAU. We thank Prof. Mike Barlow for a critical reading of an earlier
version of the paper and the referee, Prof. Manuel Peimbert, for stimulating
comments.

\end{document}